# LARGE IMPACT FEATURES ON GANYMEDE AND CALLISTO AS REVEALED BY GEOLOGICAL MAPPING AND MORPHOMETRY

Oliver L. White[a,b], Jeffrey M. Moore[b], Paul M. Schenk[c], Donald G. Korycansky[d], Andrew J. Dombard[e], Martina L. Caussi[e], and Kelsi N. Singer[f].

[a]SETI Institute, 339 Bernardo Avenue, Suite 200, Mountain View, CA 94043
[b]NASA Ames Research Center, MS 245-3, Moffett Field, CA 94035
[c]Lunar and Planetary Institute, 3600 Bay Area Blvd., Houston, TX 77058
[d]Department of Earth and Planetary Science, University of California, Santa Cruz, CA 95064
[e]Dept. of Earth and Environmental Sciences, University of Illinois at Chicago, 845 W. Taylor St., Chicago, IL 60607
[f]Southwest Research Institute, 1050 Walnut Street, Suite 300, Boulder, CO 80302

Manuscript pages: 71
Figures: 6
Tables: 3

Present address of corresponding author: Oliver White
MS 245-3
NASA Ames Research Center
Moffett Field
CA 94035
owhite@seti.org
281-309-8152 (Cell)

**Abstract.** The icy Galilean satellites are host to the broadest range of impact feature morphologies in the Solar System. Various hypotheses suggested to explain the diverse appearances of these impact features consider the effects of melt generated by the impact, the physical state of the subsurface at the time of impact, and the impactor characteristics. As part of a larger effort that uses mapping and modeling to assess the role of each of these factors in the formation and evolution of these impact features, we have performed topographic and geological mapping of 19 large impact features on Ganymede and Callisto in order to gather morphometric and crater age data that allow us to quantify how their morphologies transition with size and age. The impact features are divided into two main morphological groups, specifically craters (subdivided into pit, dome, and anomalous dome craters), and penepalimpsests/palimpsests, with these latter two forming end members of a single group. The transitions between pit, dome, and anomalous dome craters appear to be size-dependent, but for diameters above ~170 km the morphological size-dependency between anomalous dome craters, penepalimpsests, and palimpsests breaks down. A few impact features appear to be transitional between those of anomalous dome craters and penepalimpsests, displaying characteristics of each. The morphologies of pit and dome craters appear to be independent of their age or the inferred physical state of the subsurface, indicating that the impacts that formed them were small enough to only ever penetrate into a shallow, cold, rigid ice layer. Their morphologies instead derive primarily from the freezing, expansion, and drainage of a pocket of impact melt that formed underneath them, which contributes to the development of their central pits and surrounding raised annuli, while the formation of the domes can occur via viscous relaxation of the pit topography, perhaps supplemented by extrusion of melt into the dome to explain their surface textures. The subdued rims and floors of anomalous dome craters indicate the increasing effect of a weak subsurface layer on impact feature morphology with increasing size, specifically a warm ice layer, but their prominent annuli and pits also indicate that mobilization of impact melt is still an important factor. The very low topographic relief of penepalimpsests and palimpsests indicates that these impacts penetrated the ice shell to liberate and mobilize very large volumes of pre-existing liquid from a subsurface layer, with little contribution to the final feature morphology from impact melt. Penepalimpsests are distinguished from palimpsests by the higher frequency of concentric ridges within their interiors, indicating a generally more robust state of the subsurface that could support the rotation and uplift of some solid material in a concentric configuration during impact, even if a crater-like depression could not be supported. The overlap of anomalous dome craters, penepalimpsests, and palimpsests in terms of diameter as well as age indicates that impactor size, time of impact, as well as location on the satellite (and hence variation in temperature gradient across the satellite) are all factors in determining which of these morphologies emerges.

## 1. Introduction

The icy Galilean satellites (Europa, Ganymede, and Callisto) are host to a variety of large impact features that are, if not unique to these bodies, rarely encountered on planetary and satellite surfaces in the Solar System. These features are morphologically diverse and include impact craters with central pits and domes that transition to larger, so-called "penepalimpsests" that themselves transition to "palimpsests", which appear as circular albedo features with negligible topographic relief and absent crater rims (Passey and Shoemaker, 1982; Moore and Malin, 1988; Schenk and Moore, 1998; Schenk et al., 2004a). The particular circumstances that lead to such morphological variety are likely governed by the interplay of several factors that may include the presence or absence of liquid water (at depth below the surface, or generated during the impact) vs. warm ice (again, either pre-existing or impact-generated) (Croft, 1983; Moore and Malin, 1988; Schenk, 1993, 2010; Senft and Stewart, 2011); the lithospheric temperature gradient (Bray et al., 2014); surface gravity (as compared to smaller gravity on mid-sized satellites, where these types of impact features are not found); and the characteristics of the impactor, specifically its size, velocity, composition, and the angle of impact. The present study forms part of a larger effort to evaluate specific and testable hypotheses about the role of each of these factors in the formation and evolution of these impact features on Ganymede and Callisto. We perform topographic and geological mapping of 19 impact features on Ganymede and Callisto that we use to quantify how the morphometric relations between mapped components of these features transition with changing size and bulk morphology. In addition, we have performed crater counts for all the mapped impact features that we use to estimate crater ages based on comparison to R-plots for young impact features and terrains on Ganymede and Callisto presented in Schenk et al. (2004) for which Zahnle et al. (2003) have derived age estimates based on modeled impact rates at the two satellites. We interpret our results in terms of their implications for how the state of the ice shell, in particular its thickness above an internal liquid layer or melt pocket (whether pre-existing or formed by the impact event), responds to the impact process for differently-sized impacts and how variation in impact feature morphology within a certain size class can indicate evolution of the ice shell over the historical courses of these satellites. By achieving as comprehensive a view as possible of the structures of these selected impact features, our maps also serve as the primary ground-truth for numerical modeling of the impact process and long-term evolution of impact features (Korycansky et al., 2022a,b; Caussi et al., 2024).

## 2. Background

The characteristics of impact features provide clues to the physical state of the target near-surface. For instance, there is a typical sequence of crater properties as a function of diameter that reflects the physics of the impact process (Melosh, 1989). The smallest craters are “simple”: bowl-shaped features in the landscape with a typical ratio of depth *d* to diameter *D* of ~1 to 6. As craters increase in size, a transition occurs to “complex” craters that are shallower in relation to their sizes, with flatter floors, and which eventually develop a central peak. Increasing size brings a transition from the central peak to a ring around the center (so-called “peak-ring” craters). The largest impacts give rise to “multi-ring” features that are hundreds to thousands of kilometers in diameter. Transitions between crater types on terrestrial targets are influenced by several factors, principally the target composition, structure, and gravity. For instance, the diameter at which the transition from simple to complex craters occurs is inversely proportional to the surface gravity of the target (Pike, 1988). The sequence described above (simple to complex to peak-ring, etc.) is most clearly visible on bodies with rocky surfaces like the Moon, Mercury, and Mars, although icy bodies (such as outer Solar-System satellites) also manifest these transitions.

However, detailed examination of the three Galilean satellites with icy surfaces (Europa, Ganymede, and Callisto) reveals a number of more complicated and unique features on the surfaces of these bodies (Schenk et al., 2004a), particularly on the latter two. First identified in Voyager observations (Passey and Shoemaker, 1982; Moore and Malin, 1988), a small number of these impact features have been imaged at better resolution by the Galileo mission. These features exhibit shallower depth-to-diameter ratios compared to craters of similar diameter on rocky bodies. They contain dome-like central structures that have few analogues on other bodies in the Solar System, including the icy satellites of Saturn, Uranus, or Neptune. We designate these formations as "Large Impact Features". The present study focuses on Ganymede and Callisto, as Europa's surface is rather young (~108 Myr) (Greeley et al., 2004, Bierhaus et al., 2009, Schenk and Turtle, 2009) and exhibits a sequence that is distinctly different from that seen on the other satellites, extending over a smaller size range (Schenk et al., 2004a). Ganymede and Callisto have many more of these features, over a much wider size range, and all of which fit into a shared classification scheme.

As described by Schenk et al. (2004a), smaller craters on Ganymede and Callisto (with diameters <35 km) follow the traditional sequence of simple to complex, with a transition at $D$ ~3 km. Compared to the familiar lunar population, simple craters have similar depth-to-diameter ratios of ~0.3, while complex craters are ~40% to ~70% shallower than lunar ones (Schenk, 1991, 2002). Craters in the size range ~35 < D < ~75 km exhibit central pits, as seen in Figure 1a. Central pit craters are rare on the Moon and Mercury, though occasionally seen on Mars.

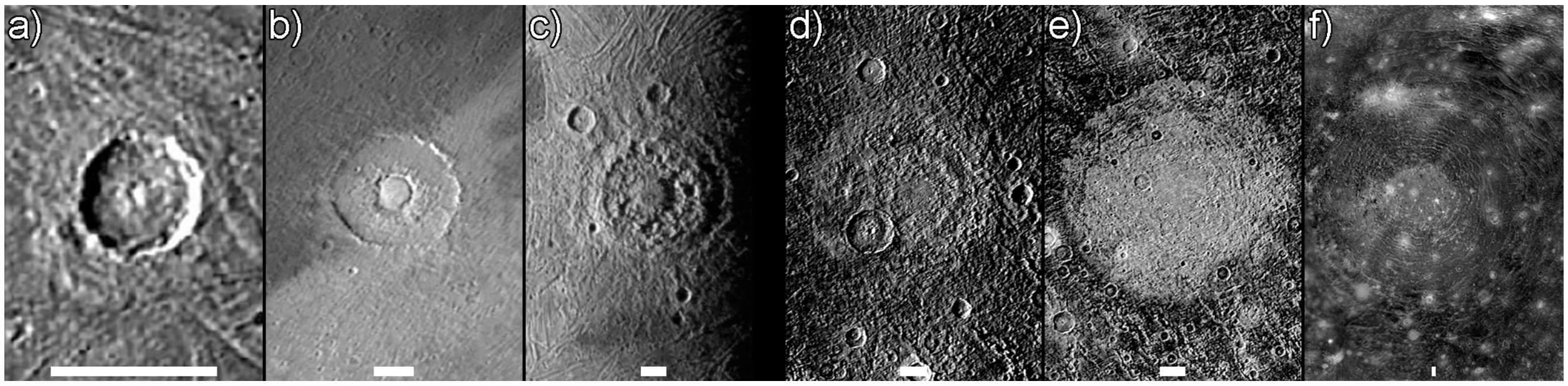

Figure 1. Morphological classification scheme for Large Impact Features on the icy Galilean satellites. The scale bar at the bottom of each panel is 40 km long. Examples shown are (a) an unnamed pit crater at 2.4°S, 132.0°W on Ganymede; (b) Enkidu, a dome crater at 26.5°S, 34.8°E on Ganymede; (c) Serapis, an anomalous dome crater at 12.2°S, 43.9°W on Ganymede; (d) Nidaba, a penepalimpsest at 17.6°N, 123.0°W on Ganymede; (e) Memphis Facula, a palimpsest at 14.2°N, 131.8°W on Ganymede; and (f) Valhalla, a multi-ring feature at 15.2°N, 55.6°W on Callisto.

More complex features are found for diameters above 75 km on Ganymede and Callisto. Five different types have been identified (Passey and Shoemaker, 1982; Schenk and Moore, 1998; Schenk et al., 2004a), over overlapping size ranges for the various types. The distinguished feature types are dome craters, anomalous dome craters, penepalimpsests, palimpsests, and multi-ring features. The sequence is shown in panels (b) to (f) of Figure 1.

Dome craters (Figure 1b) are seen over diameter ranges of ~75 to >100 km. These craters are distinguished by central domes with rounded relief up to ~1.5 km above the surroundings. They are typically 0.8 to 1.6 km in depth with low scarps at their rims. Dome diameters relative to total crater diameter typically increase with crater size. Anomalous dome craters (Figure 1c) range from >100 km to ~250 km in diameter, showing prominent, central domes surrounded by a ring of rugged massifs, but with muted rim scarps. The dome/diameter ratio of ~0.4 is constant with size. Anomalous dome craters generally seem to be much shallower than dome craters. Counts of superposed craters suggest that anomalous dome craters may be generally older than dome craters (Schenk et al., 2004a).

Penepalimpsests and palimpsests are large, circular, typically bright patches seen on Ganymede and Callisto, ranging from >150 to ~350 km in diameter. Crater rims in the traditional sense are lacking in penepalimpsests (Figure 1d), but low, nested sets of arcuate ridges can be seen in low-angle illuminated images. About 45 features have been classed as palimpsests (Figure 1e), being distinguished from penepalimpsests by their occurrence on older terrain and relative lack of positive relief features that are organized radially. As with dome and anomalous dome craters, the distinction between penepalimpsests and palimpsests might be relative age – palimpsests occur on older terrains than penepalimpsests (Schenk et al., 2004a).

Finally, a few formations, including Gilgamesh on Ganymede and Valhalla on Callisto, have been classified as multi-ring features, with Gilgamesh qualitatively resembling multi-ring basins like Orientale on the Moon. Valhalla (Figure 1f) and Gilgamesh are ~350 and ~590 km in diameter respectively. These multi-ring features have numerous graben and inward-facing scarps surrounding a palimpsest-like central region.

Impact formations have characteristics that are ultimately governed by the interplay of several factors: 1) the presence or absence of liquid water (pre-existing at depth below the surface, or generated during the impact) vs. warm ice (again, either pre-existing or impact-generated); 2) the lithospheric temperature gradient; 3) surface gravity (as compared to significantly smaller gravity on mid-sized satellites, where the features of interest are not found; and 4) the characteristics of the impactor, including composition and impact velocity. For the size scale of impact feature that we are interested in – the so-called "gravity regime" – the immediate consequences (such as the size of the crater formed soon after the impact) are largely governed by the impactor mass (or its diameter), the surface gravity of the target, and the impactor velocity. In hypervelocity impacts (where the impact velocity far exceeds the sound speed within the impactor or target), these characteristics are more determinant of crater features than the exact composition or material strength of the impactor, particularly for multi-km diameter impactors. The longer term (and ultimately final) evolution of a crater or other impact feature is dependent on characteristics of the target substrate: the gravity, thermal profile, composition, and material properties all play vital roles. The impact itself sets the initial conditions for the long-term evolution that unfolds on timescales ranging from hours to millions of years. Specific processes, like melting/freezing of the substrate (either rock or ice), heat transport (by conduction or sub-surface liquid transport), and viscous relaxation all have characteristic timescales.

Particularly relevant is the presence of liquid water in the sub-surface of the target. The water may be pre-existing layers at depth, or melt pools created by the deposition of energy by the impact. Post-impact flow and re-freezing potentially exert a strong influence on the ultimate shape of the craters, and melt-water/substrate evolution may be complex. In addition to the obvious process of freezing (with important effects due to heat release and expansion from the phase change), there is also the question of melt diffusion into the substrate via percolation, with concomitant mixing and/or melting of substrate ice. Zahnle et al. (2014) also raised the possibility, in connection with impact-generated lakes on Titan, of Rayleigh-Taylor instability due to melt: melt-water is denser than ice, and heat generated in the impact, or diffused from the melt, might induce sufficient reduction of the ice viscosity to enable overturn of the ice below the pool. Understanding the production and distribution of melt by impacts and its subsequent evolution and effects is likely key to explaining the appearance of Large Impact Features on the Galilean satellites.

Various hypotheses have been suggested over the last 40 years to explain the appearances of Large Impact Features (depicted in Figure 2), involving either warm ice or melt produced by the impact, or pre-existing warm ice or liquid water in the target substrate. In the first (Figure 2a), the impact feature's characteristics are determined by warm ice, which is produced in the impact on or above the surface. Little or none of the initially emplaced ice extends farther than the central regions of the impact feature. This hypothesis is similar to the idea of a "warm plug" of post-impact surface, which Senft and Stewart (2011) observed in their impact simulations. In the second (Figure 2b), suggested by Moore and Malin (1988), warm subsurface ice produced in the impact is the primary driver of impact feature evolution: an extensive lens of warm ice is emplaced and its evolution produces the feature's final form. In the third (Figure 2c), suggested by Croft (1983), liquid water produced in the impact plays a crucial role in shaping the various impact feature types: pools of melt of greater or lesser extent are created in the subsurface below each crater; on freezing, the expansion and subsequent relaxation of the features produce their forms. In the fourth (Figure 2d), suggested by Schenk (1993, 2010), the primary determinant is pre-existing, sub-surface ice at moderate depth (~10 km), the response of which drives the evolution of the various impact features. In the final hypothesis (Figure 2e), warm, near-surface ice at a depth of ~3 km, which may connect to a liquid layer at depth, is the main agent in driving the impact feature's evolution.

Pit craters have been the subject of previous studies aimed at elucidating formation mechanisms (Bray et al., 2008, 2012, 2014). These have also employed a combination of topographic and geological mapping plus impact modeling in an effort to constrain substrate properties; for instance Bray et al. (2014) were able to reach conclusions about the thickness of the coldest surface brittle-ice layer on Europa (about 7-10 km) with the concomitant subsurface temperature gradient (0.025 K $m^{-1}$), based on the results for the morphology of the craters as a function of size and layer depth. Our study applies these techniques to an expanded Large Impact Feature size range that covers all impact feature morphologies between pit craters and palimpsests.

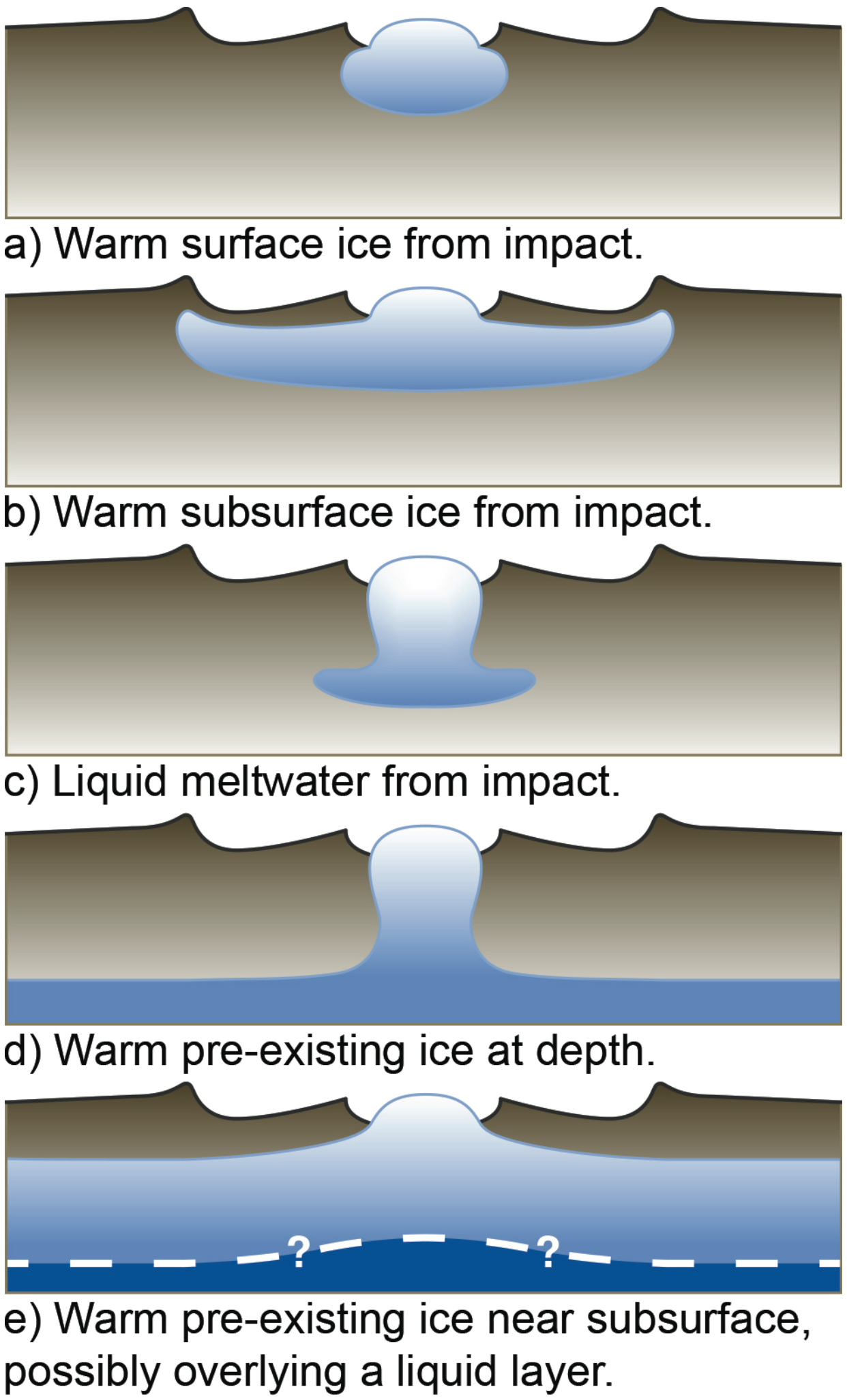

Figure 2. Different hypotheses for the initial state of a dome crater, which are described in the text. Hypotheses (a) to (c) all involve impact-produced warm ice or liquid water (melt), while hypotheses (d) and (e) involve pre-existing ice in the substrate.

## 3. Generation of Geological Maps, Digital Elevation Models, and Morphometric Statistics

Interpreting Large Impact Features is a challenge, partly due to incomplete data from the Galileo mission and Sun angles that were not always favorable for producing elevation maps. In this paper we describe the production of new digital elevation models (DEMs) and geologic maps that facilitate the interpretation of these features. We use these products to map the common facies that form our selected impact features (which vary for features in different size classes), and to obtain morphometric measurements of these facies. These statistics are then used to quantitatively compare impact feature morphology and to gauge how certain morphometric relationships of the impact features change with varying feature size. In addition, our crater age estimates of these impact features provide some indication of how impact feature morphology may not only depend upon size, but also on the state of the ice shell at the point during the satellite's history when impact occurred. Our geologic mapping establishes the real-world constraints for modeling, and so our maps and measurements form the observational basis for our hydrocode and finite element modeling, the results of which are reported in Korycansky et al. (2022a,b) and Caussi et al. (2024). Together, these endeavors offer explanations for the morphologies of these features in terms of the geological processes during and after the impact, and why there is the transition among impact feature types from pit craters through to palimpsests.

Localized mosaics assembled from Voyager and Galileo images have been generated for 19 Large Impact Features on Ganymede and Callisto (16 on Ganymede and 3 on Callisto) using the USGS ISIS3 image processing package. We use these mosaics as cartographic bases for our geologic mapping of these features, which are listed in Table 1 and are representative of five of the six impact feature morphologies shown in Figure 1. We omit the largest impact feature class, multi-ring features, from our survey as these would entail a regional-scale mapping project far in excess of the local-scale mapping that will be performed for the other classes. We specify the pixel scale of the highest resolution imaging that covers each impact feature. Where peripheral portions of the impact features (often ejecta blankets) are seen to extend beyond this high resolution imaging, we overlay this imaging onto the Ganymede and Callisto global mosaics (both projected at 1 km/pixel, although the source imaging used to make the mosaics is often coarser than this) and extend mapping into these less well-imaged areas. Where data allows, we have also generated DEMs using two techniques separately or in combination: stereo photogrammetry (Schenk et al., 1997; Schenk and Bulmer, 1998) and photoclinometry (or shape-from-shading) (Schenk, 2002). Our stereo method is an automated photogrammetry package based on scene-recognition algorithms, which match albedo patterns in finite-sized patches in each of the two stereo images, from which parallax and the corresponding difference in elevation can be determined. Our photoclinometry method derives from that described in Schenk et al. (2004b) and Schenk and Williams (2004), and involves calculating a slope for each location on the surface based on a photometric model of the brightness variation with solar elevation. The slope values are then integrated along parallel lines to produce a map of elevation differences. Where needed, high Sun angle (and often lower resolution) images are used to remove brightness variations due to intrinsic albedo variations. Photoclinometry DEMs resolve topographic features at the pixel scale of the original imaging, but they are affected by topographic undulations on regional scales. Controlling photoclinometric DEMs using coincident stereo-derived elevation data preserves high-resolution information while effectively eliminating the long-wavelength imprecision that can affect photoclinometric topographic mapping, and so provides robust topographic information at both low and high spatial frequencies (e.g. Schenk, 2002). We have generated photoclinometric DEMs for 16 of the 19 impact features and stereo DEMs for four of them. Three of these stereo DEMs have been used to control the photoclinometric DEMs that also cover the relevant impact features. Data dropouts can afflict some of the DEMs, either due to gaps in spacecraft coverage, or due to the presence of long shadows that obscure the terrain. We have co-registered our DEMs with the image mosaics used as base maps. Generation of neither stereo nor photoclinometric DEMs is viable for the three largest impact features in our survey (Zakar, Epigeus, and Memphis Facula). The base maps and (where available) DEMs for each impact feature are shown in Figure 3.

**Table 1.** The 19 impact features on Ganymede and Callisto (those on the latter indicated by asterisks) for which geological and (where possible) topographic mapping has been performed. Impact features are listed in order of increasing diameter. Stereophotogrammetric (PG) and photoclinometric (PC) DEMs have been produced for most impact features, with PG DEMs being used to control PC DEMs in three cases (PC-PG). We also list the terrain type that each impact feature superposes: either dark, cratered terrain or bright, grooved terrain (the latter specific to Ganymede).

| Name | Feature class | Center coordinates | Feature diameter (km) | Pixel scale of high resolution imaging (km) | Stereophotogrammetric (PG) or photoclinometric (PC) DEM | Terrain type superposed |
|---|---|---|---|---|---|---|
| Achelous | Pit crater | 61.9°N, 11.8°W | 40 | 0.18 | PC | Bright, grooved |
| Lugalmeslam | Pit crater | 23.7°N, 166.1°E | 64 | 0.15 | PC | Dark, cratered |
| Isis | Pit crater | 67.3°S, 158.8°E | 75 | 1 | PC-PG | Bright, grooved |
| Tindr* | Pit crater | 2.3°S, 4.5°E | 76 | 0.14 | PC | Dark, cratered |
| Eshmun | Dome crater | 17.5°S, 167.9 °E | 101 | 0.5 | PG | Bright, grooved |
| Melkart | Dome crater | 9.9°S, 173.9°E | 104 | 0.185 | PC | Dark, cratered |
| Osiris | Dome crater | 38.0°S, 166.3°W | 107 | 0.775 | PC | Bright, grooved |
| Har* | Anomalous dome crater | 3.5°S, 2.0°E | 110 | 0.14 | PC | Dark, cratered |
| Doh* | Anomalous dome crater | 30.6°N, 141.4°W | 128 (inferred) | 0.085 | PC | Dark, cratered |
| Neith | Anomalous dome crater | 29.5°N, 7.0°W | 170 | 0.14 | PC | Dark, cratered |
| Hathor | Penepalimpsest | 66.9°S, 91.3°E | 173 | 0.5 | PC-PG | Bright, grooved |
| Teshub | Palimpsest | 68.3°S, 80.7°E | 188 | 0.5 | PC-PG | Dark, cratered |
| Anzu | Anomalous dome crater | 63.5°N, 62.7°W | 193 | 2 | PC | Bright, grooved |
| Buto | Penepalimpsest | 13.2°N, 156.5°E | 235 | 0.19 | PC | Dark, cratered |
| Serapis | Anomalous dome crater | 12.4°S, 44.1°W | 253 | 2 | PC | Bright, grooved |
| Nidaba | Penepalimpsest | 17.8°N, 123.4°W | 265 | 0.8 | PC | Dark, cratered |
| Zakar | Palimpsest | 31.3°N, 26.3°E | 265 | 1.9 | - | Bright, grooved |
| Epigeus | Palimpsest | 23.0°N, 179.4°E | 349 | 0.95 | - | Bright, grooved |
| Memphis | Palimpsest | 14.1°N, 131.9°W | 354 | 0.8 | - | Dark, cratered |

Our geologic mapping follows the principles of the mapping of extraterrestrial bodies as outlined in Wilhelms (1972, 1990) and Skinner et al. (2018). We have created our maps in ArcMap, using standardized mapping conventions. Some of the assumptions of geologic mapping derived from spacecraft data are: (1) the scene contains (at least some) landforms that lend themselves to expert categorization; and (2) the landforms represent the surface manifestations of sequences of deposits or facies separated by discrete boundaries for which the rules of superposition and cross-cutting apply and are hence amenable to being stratigraphically and structurally organized. We have used the DEMs iteratively with the base maps to define facies, which constitute the major structural components of the impact features. Where the locations of the contacts separating these facies have been determined confidently, they are mapped as solid lines, but where there is some ambiguity in the exact location (which is often a consequence of low image resolution as in the cases of Isis, Eshmun, and some of the larger palimpsest impact features), the contact is marked with a dashed line.

Our geologic mapping is performed primarily to reduce these impact features to their recurring genetic elements for which morphometric statistics and relationships can be measured and compared to those of other impact features in different size classes. Linear dimensions of the impact features and their constituent landforms that we measure include the diameters of the impact features and their annuli, pits, domes, and central plains, the heights of crater rims, annuli, and domes, and the depths of crater floors and pits. "Impact feature diameter" is the rim crest-to-rim crest diameter in the case of pit, dome, and anomalous dome craters, and the diameter of the outer boundary of the undulating plains in the case of penepalimpsests and palimpsests. We make these measurements by calculating spheroidal areas for the polygons that represent the various mapped facies using ArcMap. These areas can be used to calculate mean diameters for each facies, where each facies is idealized as being circular in planform. In addition, we can use ArcMap to calculate the mean elevation of certain facies using the associated DEMs, which can then be used to calculate the vertical relief and volumes of certain facies relative to the mean elevation of other facies. These include rim heights, floor depths, and crater volumes relative to the mean elevation of the surroundings; annulus heights, pit depths, annulus volumes, and pit volumes relative to the mean elevation of the floor; and dome heights and volumes relative to the mean elevation of the pit. We define the "surroundings" as the plains to the exterior of the rim, or if the impact feature has identifiable ejecta, to the exterior of the portion of the ejecta blanket that is identifiable in the DEM. In the case of Osiris, the DEM barely covers Osiris out to its rim and does not extend beyond the immediate ejecta blanket, so the surroundings here are defined as the small portion of the ejecta blanket contained within the DEM. The crater volume measures the volume of the entire impact feature down to the level of the crater floor, and so disregards the presence of any annulus, pit, or dome at the center of the crater, which is assumed to be occupied by plains at the same mean elevation as the floor. Therefore, the ratio of the combined area of these facies to the area of the floor (for which a volume has been calculated) is found, and multiplied by the floor volume to find the corresponding volume of the portion of the impact feature occupied by these inner facies. The sum of these two volumes is the entire volume of the impact feature. A similar technique is used to calculate the pit volume in dome and anomalous dome craters, in which case the presence of the dome is disregarded.

## 4. Mapping and Morphometry Results

We have defined 14 facies across all of our geologic maps, most of which are shared between multiple impact features, but a handful of which are unique to a single feature. The resultant maps are overlain on the base maps in Figure 3 alongside the unannotated base maps and the DEMs. As presented in Figure 3, the maps are cropped closely around the rims of the impact features and so do not extend to cover the entirety of facies mapped farther outwards, specifically ejecta blankets. The complete ArcMap versions of the maps, as well as cube files of the mosaics and DEMs used to perform mapping, are contained within a zipped folder on the figshare repository, the link to which is shown in the reference list under White and Schenk (2024). The following text lists each of the facies we have defined and describes their characteristics.

*Rim:* A circular, raised ridge forming the rim of the impact feature. In some cases, particularly for smaller impact features, the rim displays a semi-continuous, sharp crest, but it more commonly appears as a ring of rugged, hilly terrain. Rims tend to be narrow and show high topographic relief for the smaller impact feature classes, but become wider and lower for anomalous dome craters (most apparently in the case of Neith). Some rims, particularly those of the largest crater class, show instances of sculpting by secondaries (e.g. Melkart, Osiris, and Neith). With the exception of Har and Doh, rims are universally seen in pit craters, dome craters, and anomalous dome craters.

*Floor:* Plains surrounded by and depressed below the rim that form the floor of the impact feature. They are generally smooth and slightly undulating when viewed at high resolution. With the exception of Har and Doh, floors are universally seen in pit craters, dome craters, and anomalous dome craters.

*Annulus:* A ring of rugged, hilly terrain, often displaying a labyrinthine network of rounded hummocks, that rises from the floors of impact features. The annulus typically slopes gently upwards from the floor and drops sharply to form cliffs that surround the central pit(s). The elevated zone defining the annulus ranges from contiguous around its entire circumference (seen at pit and dome craters) to fragmented, with valleys extending from the central pit separating the annulus into isolated segments for anomalous dome craters such as Doh and Neith. The annulus crosscuts the contact between the rim and floor. Annuli are universally seen in pit craters, dome craters, and anomalous dome craters.

*Pit:* Smooth to slightly hummocky plains surrounded by and depressed below the annulus, as well as the floor, of the impact feature. For most impact features there is a single, central pit, but for some pit craters there are multiple small, often elongate pits distributed within the annulus (as seen at Achelous and Tindr). The outer boundaries of pits abut cliffs of the surrounding annulus, and this boundary ranges from relatively smooth, as at Isis and Har, to highly angular, as seen at Doh and Neith where valleys branch from the pit and extend into the annulus. Pits are universally seen in pit craters, dome craters, and anomalous dome craters.

*Dome:* Steep-sided edifice rising from the plains of a central pit that shows an equidimensional, often square-like, planform. The profile of the edifice tends to be curved and mound-like at lower diameters, becoming more flat-topped at higher diameters. At high image resolutions, the surface often presents a vaguely lineated, "breadcrust" texture (best seen at Doh and Neith). Domes are universally seen in dome craters and anomalous dome craters, and the elevated plains seen at the center of Hathor have also been mapped as a dome.

*Bounding trough:* Forming discontinuous troughs surrounding portions of the crater rim, this facies does not present a distinct texture in even high resolution imaging, but rather is only really apparent in the DEMs. Bounding troughs are only observed for the anomalous dome craters of Neith and Anzu.

*Ejecta:* Rough terrain surrounding an impact feature that is interpreted as an ejecta blanket on account of it displaying a texture and albedo that is distinct from that of surrounding terrain, being elevated above surrounding terrain and bounded by a lobate scarp, and/or displaying a fabric that is radial to the impact feature. This facies is commonly seen surrounding pit and dome craters.

*Impact crater:* Craters formed by subsequent impacts that are superimposed on the impact feature. Crater rims and floors are mapped as this facies, as well as ejecta where the ejecta has a distinct morphologic and topographic signature as seen in the imaging and the DEM. At Epigeus, several superimposed impact craters reaching tens of km across have associated bright and diffuse ejecta blankets, but these have been mapped with a crosshatched overlay symbol rather than as the impact crater facies. For each impact feature we map all craters above a certain minimum diameter as this facies. This diameter depends on the pixel scale of the imaging (the smallest mapped craters are always at least 5 pixels across) as well as the size of the impact feature (e.g. the smallest mapped craters at Neith are at least 14 pixels across, owing to the high resolution of the imaging combined with the large impact feature size). We use these mapped craters for our crater age statistics, and the minimum mapped crater diameters and the total number of mapped craters above this diameter are listed for each impact feature in Table 3. Impact craters superimpose most impact features across all size classes. No superimposed craters are mapped for Isis and Anzu (likely owing to low resolution imaging) and Osiris (likely owing to its very young age as implied by its very bright and prominent ejecta rays which superpose all terrains in contact).

*Outer platform:* This facies is unique to Har amongst the impact features we have mapped, and forms rugged, hilly terrain surrounding the annulus. It is elevated above surrounding terrain, and a bounding scarp is commonly visible where it is covered by high resolution imaging. Schenk et al. (2004) note that annular plateaus or pedestals surround many impact craters on Ganymede and Callisto, and suggest that they may have formed by radial flow or plastic deformation of a thicker inner portion of an ejecta deposit (e.g. Horner and Greeley, 1982; Moore et al., 1998), forming a convex snout as flow or creep halted. However, unlike these pedestals, which surround the rim of a crater, this facies is located where the crater floor and rim would normally be expected, but these individual facies cannot be distinguished. If the outer platform does represent both the floor and rim, then the individual facies may have been obscured due to having been covered by ejecta and impacted by secondaries from the neighboring Tindr impact. In addition, massive sublimation erosion unique to Callisto (e.g. Moore et al., 1999; White et al., 2016) may play a role in the absence of a rim.

*Crater chain:* This facies is unique to Har, and consists of linear chains of impact craters reaching 50 km long that are radial to the neighboring Tindr impact feature and are secondaries stemming from that impact. Individual craters within these chains cannot be distinguished, so they are mapped collectively and are not included in the crater age analysis. We note, however, that many of the individual craters mapped as the impact crater facies, and which are included in the crater age analysis, are likely also Tindr secondaries.

*Central plains:* Expanse of flat plains reaching up to 100 km across that exists at the centers of most penepalimpsests and palimpsests. The plains are generally smooth, but can display a slightly knobby texture at high resolution (as seen at Buto Facula), reminiscent of “small chaos” on Europa. They are separated from the surrounding, higher-relief undulating plains by a low, inward-facing scarp, and superpose the undulating plains and concentric ridges.

*Concentric ridges:* Arcuate ranges of low hills reaching several tens of kilometers long, which are seen in all penepalimpsests and palimpsests, with the exception of Memphis Facula. They are situated within undulating plains and together form concentric rings at varying distances from the center of the impact feature. The ridges can be common and densely spaced, as at Buto Facula or Nidaba, or infrequent and sparse, as at Teshub and Epigeus.

*Undulating plains:* This facies areally dominates the penepalimpsests and palimpsests and consists of plains that gently undulate with amplitude of a few hundred meters. They extend from the outer, scarp-defined boundary of the central plains to (in most cases) beyond the concentric ridges by a few tens of kilometers. The boundary between the undulating plains and the surrounding terrain is frequently ambiguous and is inferred based on a transition in the albedo as seen in the imaging (most apparent in the case of Memphis Facula) and surface texture as seen in the imaging or DEMs (when available); there is no distinct "rim" to the undulating plains.

*Dark, undulating plains:* This facies is unique to Memphis Facula. The majority of this impact feature is composed of regular undulating plains with an albedo that is high compared to the surrounding terrain, but towards the margins of the feature (typically within ~50 km of its boundary), portions of these plains display a lower albedo, but which is nevertheless still slightly higher that that of the surrounding terrain. Where they form large expanses, these dark, undulating plains appear to display somewhat less surface contrast at a scale of several kilometers compared to the regular undulating plains, and exhibit a vaguely concentric configuration.

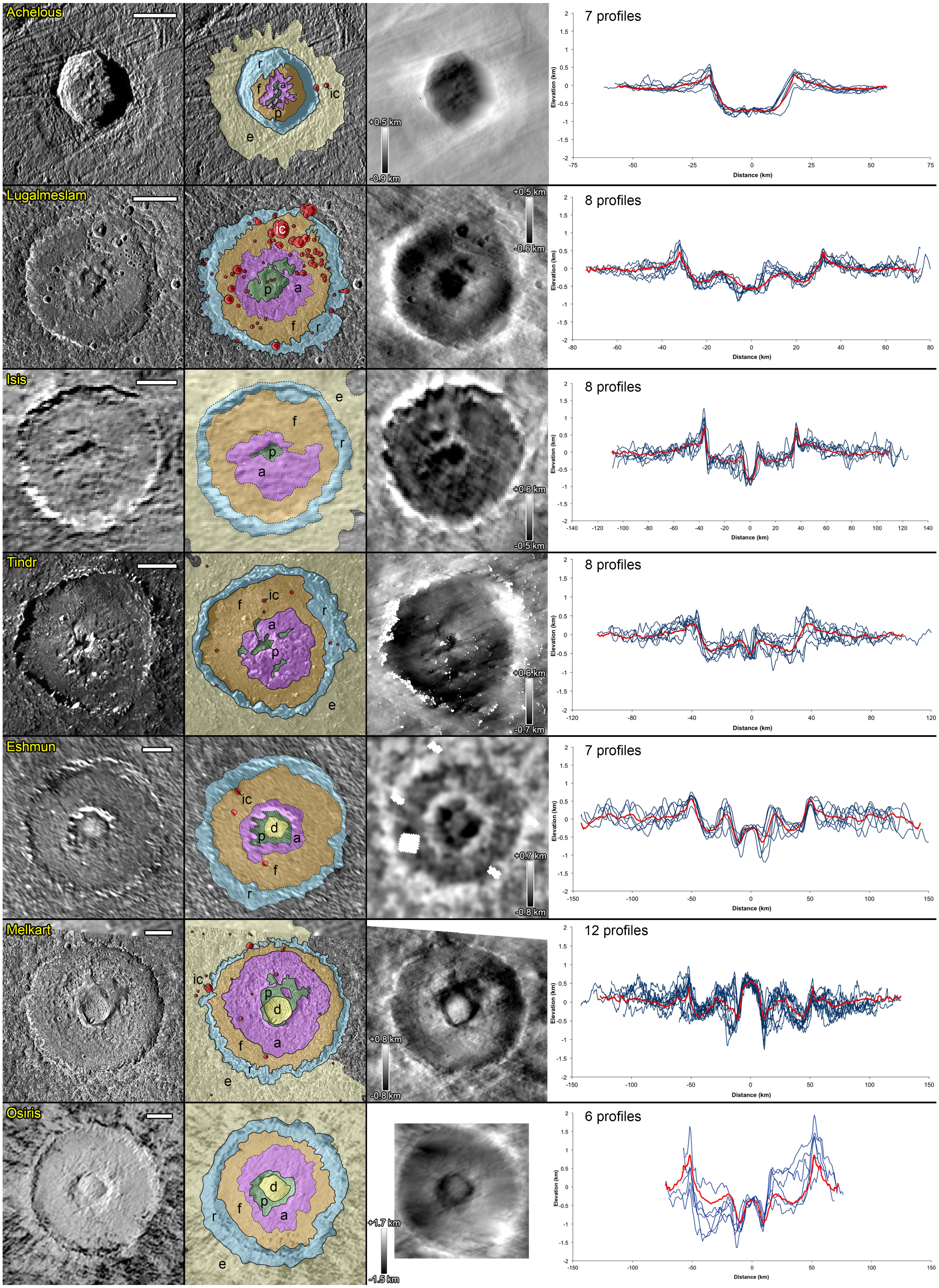

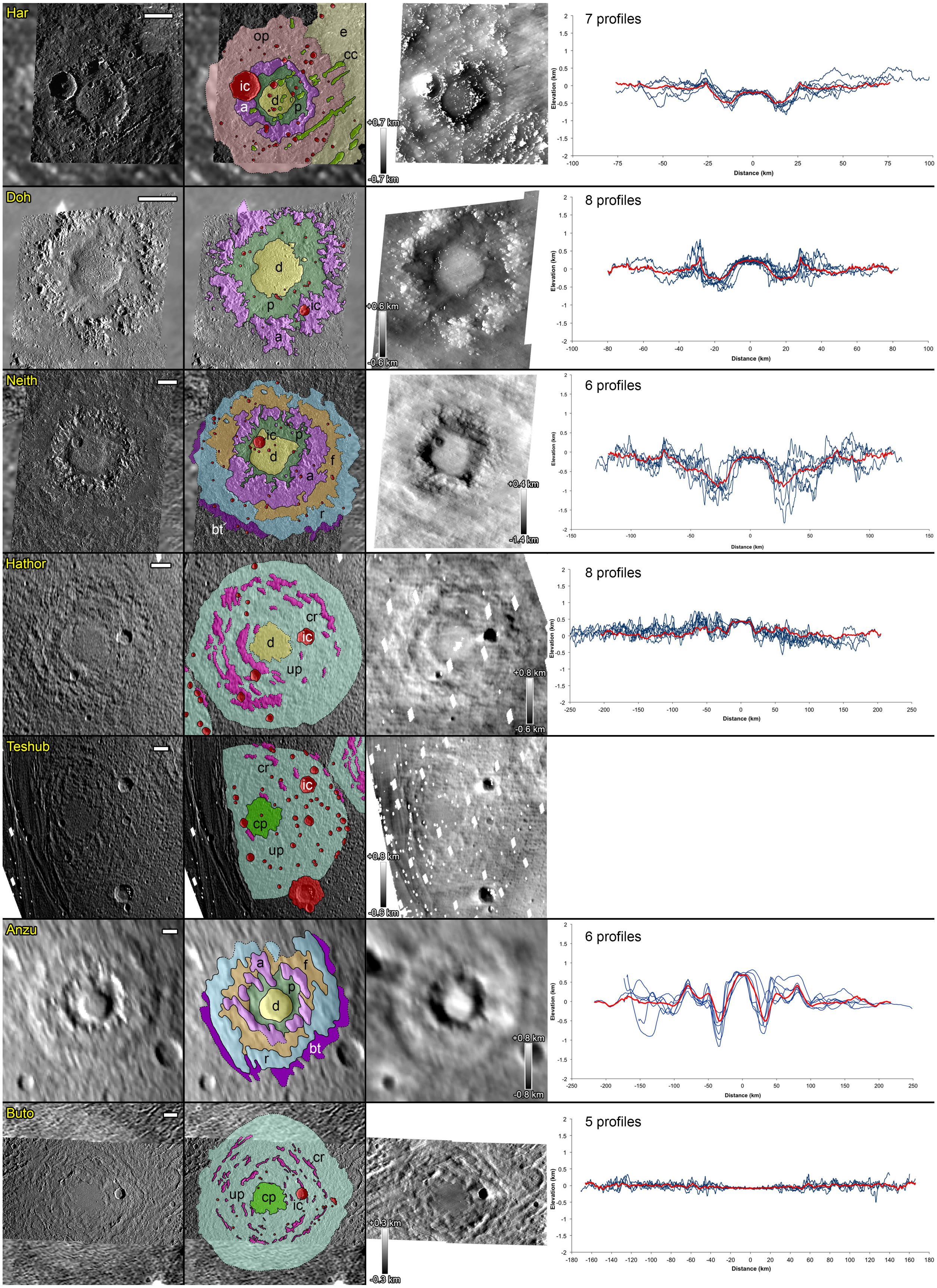

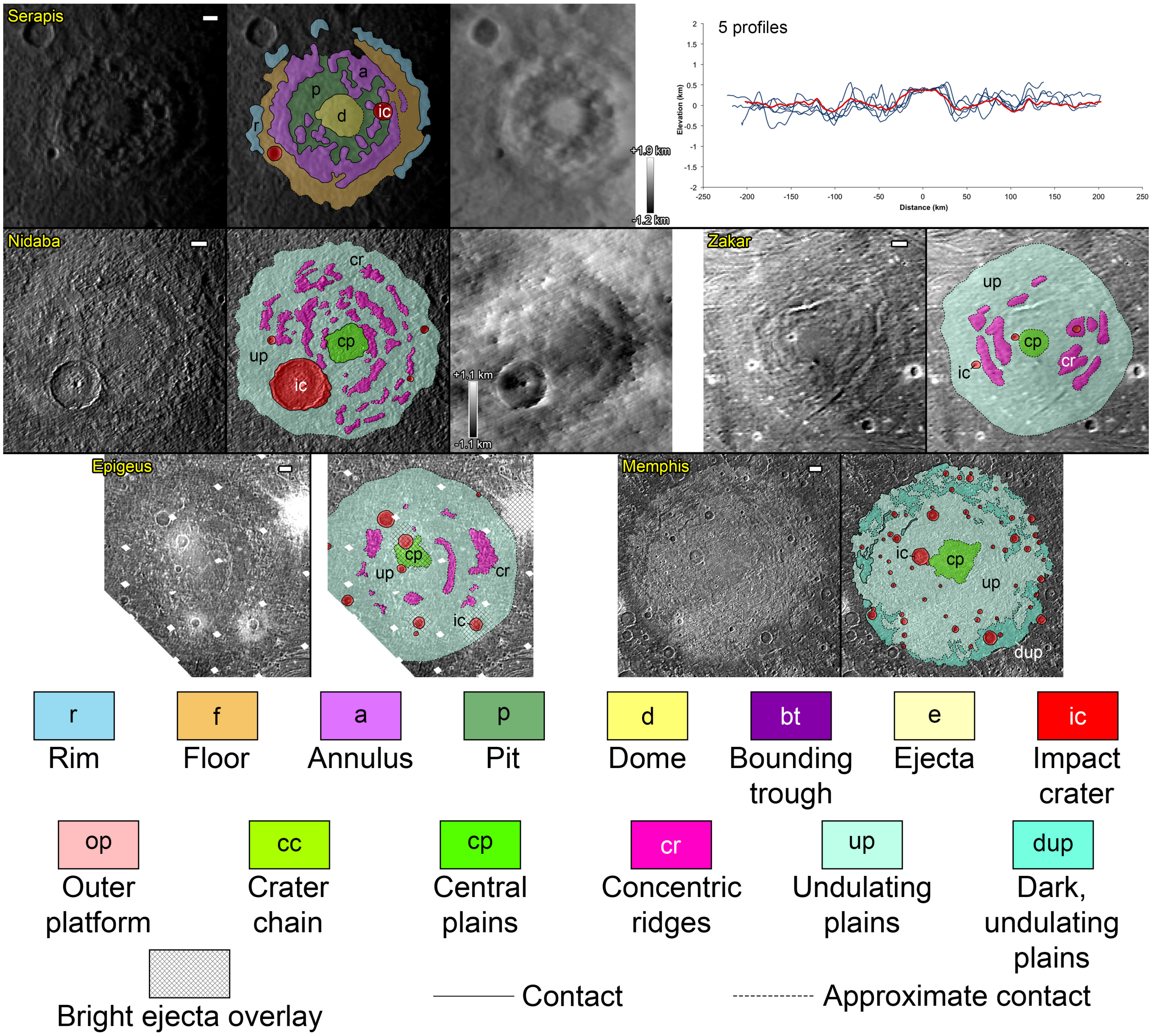

Figure 3. Maps of the 19 large impact features on Ganymede and Callisto, arranged in order of feature diameter. For each one, at left is the imaging used as the base map for geologic mapping, at center is the geologic map overlain on the base map, and at right is the corresponding DEM (available in most cases, but not for Zakar, Epigeus, and Memphis Facula), accompanied by an elevation scale. Scale bar at the top right of each base map measures 20 km. For 14 of the impact features with associated DEMs, multiple topographic profiles across the feature have been drawn (blue lines) and used to generate an average profile (red line). Profiles are all shown at the same elevation scale.

To illustrate the transition in impact feature morphology with increasing feature size, we have taken multiple (at least 5 in each case) profiles across each impact feature for the majority of those that have DEM coverage, and used these to produce a single averaged, symmetrical, 'idealized' profile for each. The raw and averaged profiles are shown in Figure 3, along with the number of raw profiles that were used to make the averaged profile. We have declined to take profiles across Teshub and Nidaba, even though they have DEM coverage, as about a third of Teshub has been obliterated by Bubastis Sulci, and the photoclinometry DEM of Nidaba shows too much fluctuation across regional scales, presumably due to variations in intrinsic albedo. A high quality DEM is available for Buto Facula, and we regard the averaged profile that we have obtained for this feature to also be broadly representative of those of Teshub and Nidaba.

We present morphometric statistics measured for each impact feature in Table 2, and plots that show how these statistics vary with increasing impact feature diameter in Figure 4. The quantities plotted against impact feature diameter include facies diameter, the ratio of facies diameter to impact feature diameter, facies relief, and facies volume. For each facies in each plot, we have approximated the relationship of the quantity to the impact feature diameter with trend lines.

**Table 2.** Morphometric statistics measured for the 19 impact features, which are displayed graphically in Figure 5. Diameters are mean values determined from the measured areas of the facies. Depths measure the difference between the mean elevation of the crater floor and pit floor beneath the mean elevation of the surroundings and crater floor respectively. Heights measure the maximum elevation of the rim, annulus, and dome above the surroundings, crater floor, and pit floor respectively. Crater, annulus, pit, and dome volumes are measured relative to the mean elevation of the surroundings, crater floor, crater floor, and pit floor respectively.

| Name | Impact feature diameter (km) | Annulus diameter (km) | Pit diameter (km) | Dome diameter (km) | Central plains diameter (km) | Rim height (km) | Floor depth (km) | Annulus height (km) | Pit depth (km) | Dome height (km) | Crater volume ($km^3$) | Annulus volume ($km^3$) | Pit volume ($km^3$) | Dome volume ($km^3$) |
|---|---|---|---|---|---|---|---|---|---|---|---|---|---|---|
| Achelous | 40 | 15 | 6 | - | - | 0.78 | 0.47 | 0.05 | 0.18 | - | 307 | 0.05 | 4 | - |
| Lugalmeslam | 64 | 34 | 18 | - | - | 0.81 | 0.29 | 0.64 | 0.17 | - | 720 | 100 | 46 | - |
| Isis | 75 | 36 | 12 | - | - | 1.44 | 0.14 | 0.71 | 0.29 | - | 538 | 43 | 36 | - |
| Tindr | 76 | 38 | 12 | - | - | 1.05 | 0.32 | 0.74 | 0.07 | - | 1002 | 123 | 9 | - |
| Eshmun | 101 | 43 | 26 | 15 | - | 0.83 | 0.23 | 0.85 | 0.40 | 0.44 | 1567 | 197 | 235 | 39 |
| Melkart | 104 | 76 | 32 | 21 | - | 1.17 | 0.30 | 1.24 | 0.21 | 1.36 | 2227 | 1069 | 226 | 226 |
| Osiris | 107 | 63 | 33 | 21 | - | 2.07 | 0.34 | 1.00 | 0.50 | 0.63 | 3076 | 426 | 407 | 57 |
| Har | 110 | 56 | 39 | 25 | - | 0.96 | - | 0.52 | 0.48 | 0.47 | - | 46 | 560 | 87 |
| Doh | 128 (inferred) | 71 | 52 | 28 | - | - | - | 0.97 | 0.16 | 0.77 | - | 314 | 336 | 182 |
| Neith | 170 | 108 | 69 | 45 | - | 0.96 | 0.05 | 0.96 | 0.70 | 1.15 | 1372 | 288 | 2556 | 952 |
| Hathor | 173 | - | - | 40 | - | - | - | - | - | 0.64 | - | - | - | 338 |
| Teshub | 188 | - | - | - | 51 | - | - | - | - | - | - | - | - | - |
| Anzu | 193 | 109 | 75 | 47 | - | 0.85 | -0.09 | 0.80 | 0.47 | 1.26 | - | 559 | 1814 | 1242 |
| Buto | 235 | - | - | - | 50 | - | - | - | - | - | - | - | - | - |
| Serapis | 253 | 180 | 140 | 60 | - | 0.67 | -0.05 | 0.74 | 0.13 | 0.73 | - | 2174 | 1336 | 1003 |
| Nidaba | 265 | - | - | - | 52 | - | - | - | - | - | - | - | - | - |
| Zakar | 265 | - | - | - | 40 | - | - | - | - | - | - | - | - | - |
| Epigeus | 349 | - | - | - | 56 | - | - | - | - | - | - | - | - | - |
| Memphis | 354 | - | - | - | 76 | - | - | - | - | - | - | - | - | - |

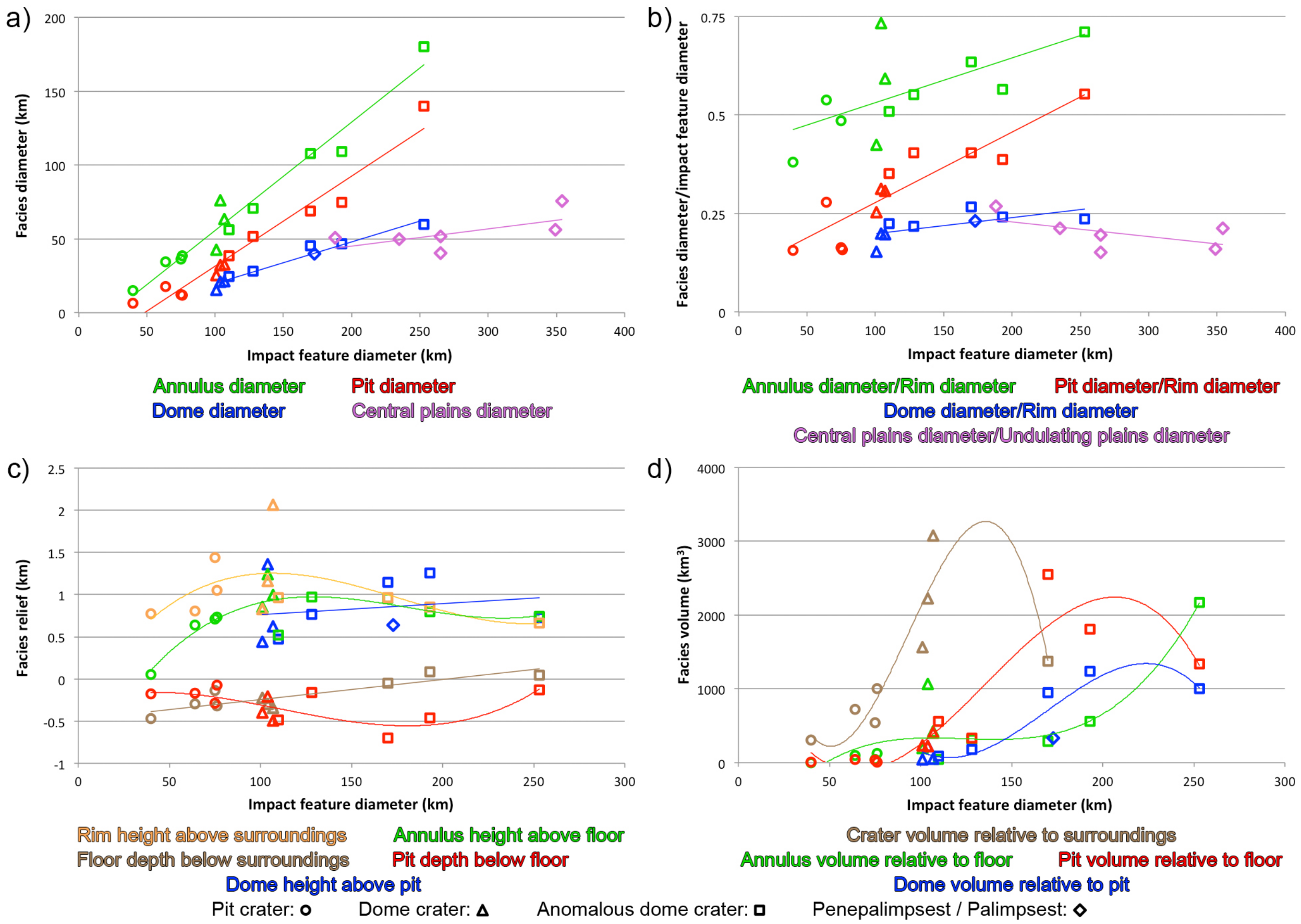

Figure 4. Morphometric statistics measured for each mapped facies as listed in Table 2, plotted against impact feature diameter. Different facies are represented by different colors, with the specific measured quantity in each plot defined by the corresponding colored text beneath. Different impact feature classes are indicated by different symbols as defined in the legend at bottom. Trend lines are defined for each facies in order to approximate the relationship of each quantity to the impact feature diameter, and are either linear or third order polynomial.

## 5. Crater Statistics and Ages

While size is clearly an important factor in determining an impact feature's morphology, gauging the relative ages of our impact features allows us to assess whether the timing of an impact feature's formation correlates with its morphology, which in turn might illuminate how the thermal and physical state of the ice shells of these satellites have changed over time. Determining ages for these impact features relies primarily on counting the craters superposed on each one, and using these counts, in combination with the total areas of the impact features as calculated using our mapping, to derive the crater spatial densities for each one. As noted in section 4, our mapping of the impact crater facies is used to derive crater counts for each impact feature. In nearly all cases this facies is mapped such that it covers a single superposing crater out to its rim, so the mean rim-to-rim diameter of the crater can be calculated based on its area. Exceptions to this routine are where a single polygon of the impact crater facies covers multiple superposing, adjacent craters, and where the impact crater facies also includes any ejecta that surround a superposing impact crater, in which case we measure the rim-to-rim diameters of each individual crater manually in ArcMap. For those impact features for which there is a steep decline in resolution from the high-resolution imaging that covers them to the surrounding global mosaic, we only perform crater counting for the portions of the impact features that are contained within the high resolution imaging. Our crater counts include craters superposed on any ejecta blanket that an impact feature may possess.

Table 3 presents crater count statistics for the 19 impact features, including the slopes of the cumulative and differential size-frequency distribution (SFD) plots shown in Figure 5. The slopes were fit by linear regression to either the cumulative or differential data points. Of the 16 impact features for which superimposed craters were observed, less than five craters were counted for Achelous, Eshmun, Serapis, Nadaba, and Zakar, and those few that were counted typically cover a narrow diameter range of a few km or less (Nidaba being a notable exception). These plots include potential secondary craters, the effect of which we will discuss on an individual basis in the following section. The differential slopes shown in Table 3 correspond to the plots exactly as shown in Figure 5b, with the exception of those for Melkart, Har, Doh, Neith, Teshub, and Memphis Facula. The data points for the lowest crater diameter ranges within the differential SFD plots for each of these impact features plot much lower than the trend defined by the other data points. This is a consequence of only a fraction of the craters within the smallest diameter range being identified due to the range overlapping with the resolution limit of the imaging, so for these impact features we discard the first data point (the smallest diameter bin) of their differential plots in order to obtain more accurate values for their differential slopes.

**Table 3.** Crater count statistics for the 19 impact features. No superimposed impact craters above the minimum diameter were observed for Isis, Osiris, and Anzu. Impact features with more than 1 and fewer than 5 superimposed craters that are above the minimum diameter are indicated in italics. Cumulative and differential slopes are for the size-frequency distributions in Figs. 6a and 6b. The distribution for Achelous is only a single point in the differential plot, so no differential slope is shown for it.

| Name | Minimum diameter of counted craters (km) | Crater count | Crater diameter range (km) | Crater counting area ($\times 10^3$ km$^2$) | Cumulative slope | Differential slope |
|---|---|---|---|---|---|---|
| *Achelous* | *1.27* | *2* | *2.22 - 2.75* | *3.14* | *-3.28* | - |
| Lugalmeslam | 1.06 | 73 | 1.17 - 8.07 | 3.82 | -2.11 | -2.55 |
| Isis | 7.04 | 0 | - | - | - | - |
| Tindr | 1.27 | 17 | 1.27 - 3.55 | 20.59 | -1.99 | -0.95 |
| *Eshmun* | *3.52* | *3* | *3.94 - 4.33* | *8.72* | *-10.42* | *-3.00* |
| Melkart | 1.17 | 47 | 1.17 - 6.23 | 38.58 | -2.47 | -3.26 |
| Osiris | 5.46 | 0 | - | - | - | - |
| Har | 1.28 | 77 | 1.32 - 21.47 | 7.77 | -2.02 | -2.41 |
| Doh | 0.89 | 20 | 0.91 - 5.65 | 3.89 | -1.86 | -2.55 |
| Neith | 1.91 | 46 | 1.91 - 12.92 | 22.28 | -2.58 | -2.84 |
| Hathor | 3.52 | 9 | 4.19 - 17.42 | 25.04 | -1.41 | -1.80 |
| Teshub | 3.52 | 50 | 3.81 - 27.39 | 29.20 | -2.17 | -2.52 |
| Anzu | 14.1 | 0 | - | - | - | - |
| Buto | 2.50 | 29 | 2.50 - 17.66 | 32.38 | -2.02 | -2.28 |
| *Serapis* | *14.1* | *2* | *20.48 - 25.68* | *43.30* | *-3.07* | *-1.00* |
| *Nidaba* | *5.63* | *4* | *8.12 - 71.37* | *55.27* | *-0.55* | *-1.78* |
| *Zakar* | *10.12* | *3* | *10.12 - 11.78* | *53.76* | *-7.11* | *-3.00* |
| Epigeus | 6.69 | 9 | 7.74 - 29.56 | 86.75 | -1.29 | -0.90 |
| Memphis | 5.26 | 55 | 5.26 - 29.62 | 94.25 | -2.33 | -2.99 |

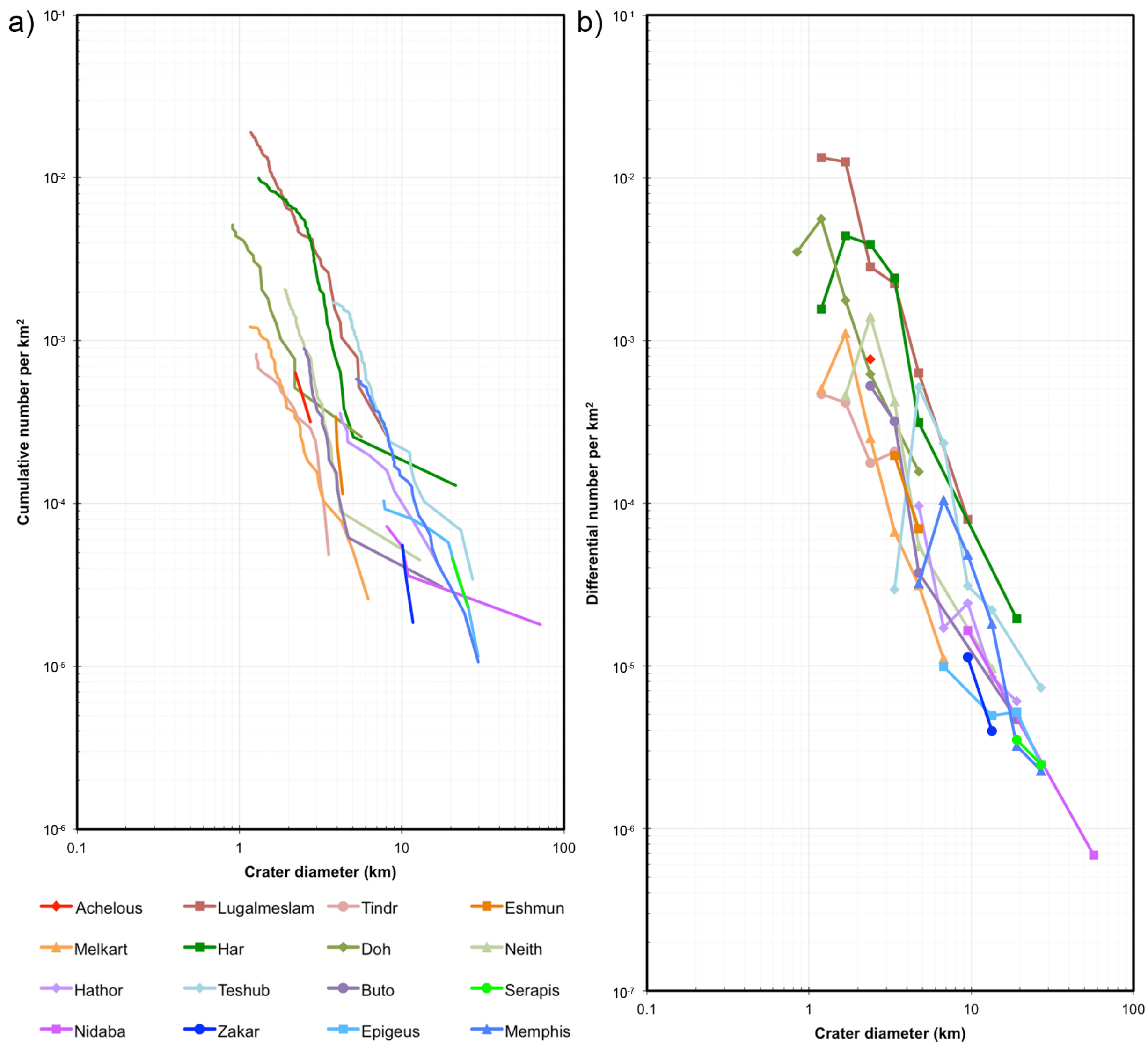

Figure 5. Cumulative (a) and differential (b) crater SFD plots for the 16 impact features for which superimposed impact craters were observed. The cumulative SFD plot contains data points for each individual counted crater is shown as a line plot for clarity; each crater diameter bin in the differential SFD plot is indicated with a marker.

Figure 6 shows the crater densities for the 16 impact features individually in the relative or R-value format, in which a differential crater density represented by a power law with a slope of $q$ is normalized by a size distribution with $q = -3$ ($dN/dD \propto D^q/D^{-3}$) (Crater Analysis Techniques

Working Group, 1979). Error bars on the points are the unit R-values divided by $\sqrt{N}$, where $N$ is the number of craters mapped in that unit. We have omitted the R-value data points for the smallest crater diameter ranges of the aforementioned impact features that display abnormally low values. We also group the R-plots together according to parent satellite, and juxtapose them with R-plots created by Schenk et al. (2004) for young surface features (including the Gilgamesh multi-ring feature and the bright terrain on Ganymede, and the Lofn crater and Valhalla multi-ring feature on Callisto), as well as for nearly global crater counts on both satellites that consider craters larger than 30 km and 50 km diameter on Ganymede and Callisto respectively. Zahnle et al. (2003) have estimated the age of Ganymede's bright terrain to be ~2 Ga, and that of Gilgamesh to be 700 Ma to 1 Ga depending on whether the impact feature postdates or predates nonsynchronicity. Zahnle et al. (2003) have also assigned nominal ages of ~2 Ga to both Lofn and Valhalla on Callisto, although the young age is very uncertain for Valhalla, which is more densely cratered. Because Valhalla is near the apex of motion on Callisto's leading hemisphere, cratering rates are high so that dense cratering does not in itself imply great age (Schenk et al., 2004).

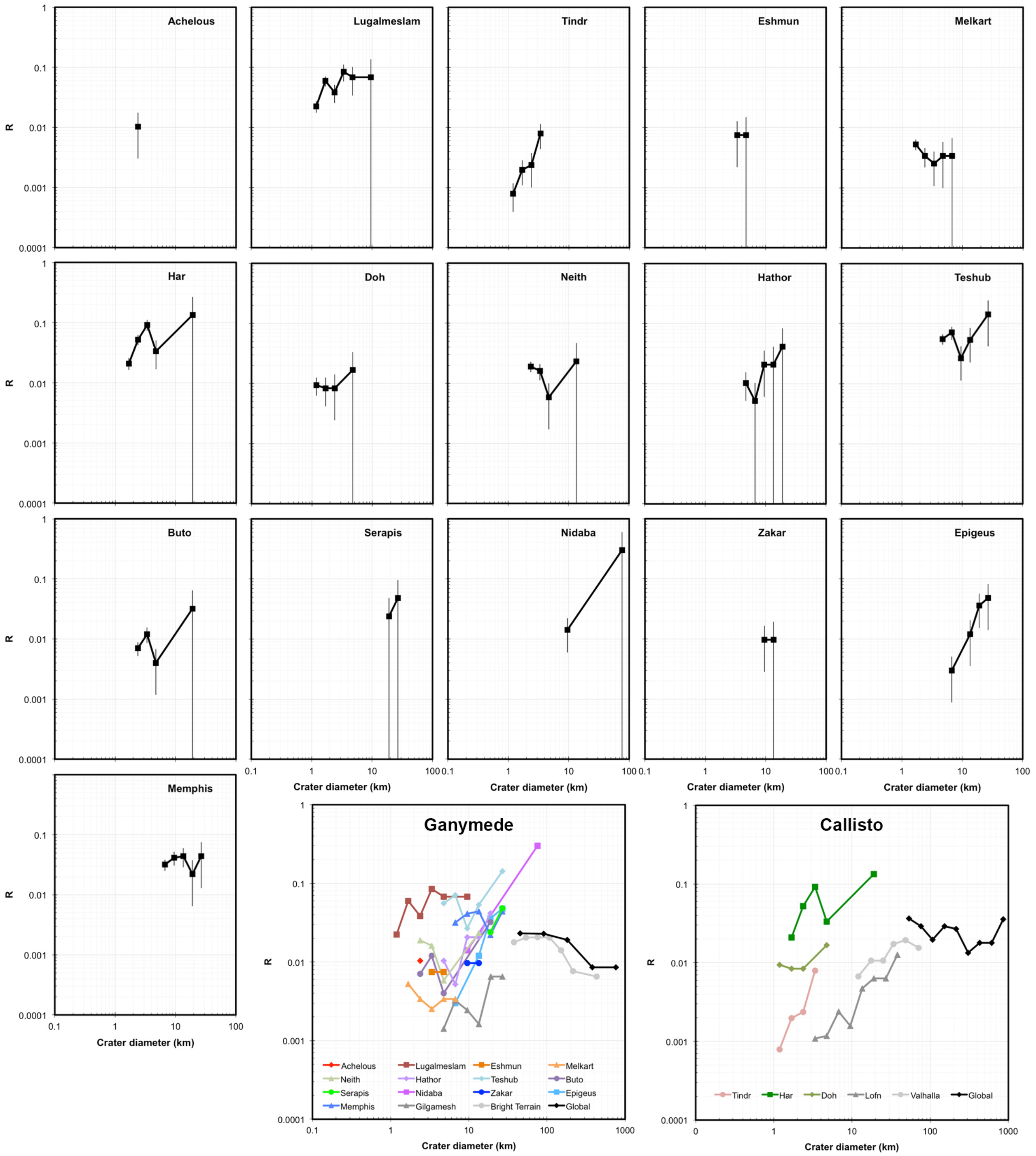

Figure 6. Individual R-plots, with error bars, for the 16 impact features for which superimposed impact craters were observed. Collective plots for impact features on Ganymede and Callisto are shown at the bottom, with different colored plots for the different impact features. The gray and black R-plots are for crater counts of young terrains and features on Ganymede and Callisto, as well as for nearly global crater counts, as presented in Schenk et al. (2004).

## 6. Discussion

In this section we discuss our impact feature mapping, morphometry, and crater statistics results in the context of how they vary across morphological classes, and the implications for the near-surface conditions that prevailed during and after their formation, referring to modeling results of Caussi et al. (2024) and Korycansky et al. (2022a,b) where relevant. We first describe mapping and morphometric trends across the full range of morphological classes before focusing on impact features within individual morphological classes in order of increasing size and what factors (including age and subsurface conditions) may contribute to morphological variety seen within each class.

All features belonging to the crater morphological classes (which include pit craters, dome craters, and anomalous dome craters) show a common configuration of a central pit (or in the case of features like Achelous and Tindr, multiple small pits) that is surrounded by a hummocky annulus, which for larger features (like Neith) develops into a system of radial spurs separated by valleys extending outwards from the pit. The annulus is surrounded by an expanse of smooth plains corresponding to the crater floor, which is in turn surrounded by an outer rim, which tends to be quite narrow and high relief for pit and dome craters, but becomes much broader and muted for anomalous dome craters. Penepalimpsests and palimpsests show a different configuration of a central expanse of smooth plains surrounded by undulating plains in which exist concentric ridges of varying spatial density. Figure 4a shows that facies diameter increases fairly linearly with impact feature diameter, but with the slope of the linear trend decreasing in the order of annulus, pit, dome, and central plains. In fact, while the ratio of facies diameter to impact feature diameter scales positively with impact feature diameter for annuli, pits, and domes, as seen in Figure 4b, this ratio scales slightly negatively for central plains. These trends indicate that as the diameter of undulating plains of penepalimpsests and palimpsests increases, there is little corresponding increase in the diameter of the central plains, in contrast to the other three facies. Domes start to appear on the floor of the central pit as the ratio of pit diameter to rim diameter exceeds ~0.25, although Lugalmeslam stands in contrast to the other three pit craters with an anomalously high pit/rim diameter ratio of ~0.28.

Figure 4c highlights the more subdued relief of anomalous dome craters relative to smaller features: pit and dome craters exhibit a trend of increasing relief of rim height, annulus height, and pit depth with increasing feature diameter, but for rim and annulus heights, this trend reaches a maximum within the dome class before decreasing with increasing feature diameter for anomalous dome craters. Pit depth increases with increasing feature diameter (excepting Doh's anomalously shallow pit) as far as the diameter of Neith (170 km), above which pit depth decreases. This is in contrast to floor depth, which decreases steadily with increasing diameter across all morphological classes (eventually culminating in Anzu and Serapis actually displaying mean floor elevations that are elevated above the surrounding terrain). Thus, while pits are generally shallower than the craters in which they are situated for pit craters, and approximately the same depth as the craters for dome craters, they are actually deeper than the craters for anomalous dome craters. There is a slight positive trend of dome height with feature diameter, with Melkart's anomalously tall dome bucking the trend. The facies that form penepalimpsests and palimpsests display little topographic relief relative to the terrain that surrounds the impact features; the only one that has any prominence is the concentric ridges, which can show relief of up to 0.5 km above the adjacent undulating plains.

Figure 4d shows that both crater and pit volume increase with impact feature diameter across the pit and dome feature classes, sharply so in the case of crater volume. These trends drop off for the larger features in the anomalous dome feature class, although we note that crater volumes are not calculated for Har and Doh (as crater floors are not identified for either of these features), nor Anzu and Serapis (as both of these features display mean floor elevations that are above the surroundings). For pit craters that show several small pits distributed within the annulus (Achelous and Tindr), the ratio of pit volume to crater volume is small (less than 0.014), increasing to ~0.065 for those pit craters with a single pit surrounded by the annulus (Lugalmeslam and Isis), and then increasing further with the transition to dome craters (0.1 to 0.14). All pit and dome craters display smaller pit volumes than crater volumes, but Neith, which displays the deepest pit and largest pit volume of any impact feature, is unique in that its pit volume is larger than its crater volume by a factor of almost 2. Dome volumes increase with increasing impact feature diameter but appear to tail off for the larger anomalous dome craters, while the elevated central zone of Hathor (which we have mapped as the dome facies) has an anomalously low volume. The ratio of dome volume to pit volume tends to be towards the lower end of the range of 0.14 to 0.75, but Melkart's dome is remarkable in that its great height (rising 1.36 km above its base compared to the mean of 0.85 km) means that it has essentially the same volume as its pit. Annulus volume shows little increase with impact feature diameter up to and including Neith, above which it shows a dramatic increase for the largest anomalous dome craters, mostly on account of their much larger areas rather than showing particularly high topographic relief. An exception to this trend is the very high volume of Melkart's annulus on account of its unusually high relief and wide diameter. Since it is the smallest impact feature that we have measured, it makes sense that Achelous also has the smallest annulus, but it is anomalously small in terms of both height (~50 m) and volume (0.05 $km^3$), giving Achelous a nearly flat floor as seen in the profile in Figure 3.

*Pit craters:* Three of the four pit craters that we have examined are on Ganymede (Achelous, Lugalmeslam, and Isis), with the fourth (Tindr) being on Callisto. Achelous and Tindr are both well resolved in imaging and topography and display few, small craters within their rims and on their ejecta blankets, with consequentially low to middling values in their R-plots (Figure 6). The dense fields of secondary craters that surround them are also indicative of their relative youth. Singer et al. (2013) mapped 630 secondaries beyond the ejecta blanket of Achelous, the largest of which is 2.7 km across. It is likely that the majority of the ~2 km-sized craters on Gula, a central peak crater that is the same diameter as Achelous and located just 70 km from it, but which has a more degraded appearance, are also secondaries from Achelous. High resolution imaging does not extend beyond Tindr's ejecta blanket as it does for Achelous's, but the majority of the ~2 km-sized craters that we have mapped on Har, located just ~30 km southwest of Tindr, are Tindr secondaries, as are the northeast-southwest-oriented crater chains that mark Har. Based on their R-plots, Tindr may be roughly contemporary with Lofn at ~2 Ga (Zahnle et al., 2003). Isis, which displays a single large pit surrounded by the annulus, is the least well-resolved pit crater, with no craters larger than its resolution limit of 7 km having been mapped. This size is about the same as the largest superposing crater that we have mapped on any of the pit craters (8 km for Lugalmeslam), and so Isis may not be as young as it seems based on its total lack of craters. However, it does exist on bright terrain, which suggests that it may date to Ganymede's middle age, as Zahnle et al. (2003) assigned an age of ~2 Ga to the bright terrain. Lugalmeslam is the most cratered of all the pit craters, and is amongst the highest plotting impact features on the R-plot. Several of its craters that are 2 km across or less occur adjacent to each other in clusters or doublets, and so may be secondaries, although the source impact for them is uncertain as there are no large impacts within the vicinity of Lugalmeslam. However, 12 of Lugalmeslam's 73 craters are larger than a few km across, the impact feature as a whole has a quite degraded appearance, it has formed within Ganymede's dark, cratered terrain, and its northeastern portion has been deformed by Mashu Sulcus, all of which point to it being the oldest of the four pit craters, likely older than the ~2 Ga isochron. Like Isis, Lugalmeslam displays a single central pit, but which has a complex planform, with a narrow spur of annulus material nearly splitting the pit in two.

The diameter range of the pit craters in our survey is quite narrow, from 40 to 76 km, and those pit craters that show multiple small, irregular pits distributed within the annulus (Achelous and Tindr) are at opposite ends of this range. This would suggest that the development of a single large pit as opposed to multiple small pits is not dependent solely on the diameter of a pit crater. We have determined that the two pit craters showing multiple small pits are younger than Lugalmeslam, the well-resolved pit crater that shows a single large pit, while Isis, the least well-resolved pit crater that also shows a single large pit, is of uncertain age based on crater statistics. Lugalmeslam and Tindr both exist within dark, cratered terrain, while Achelous and Isis both superpose bright, grooved terrain.

The modeling of Korycansky et al. (2022b) has simulated an impactor population with characteristics and parameters suitable for Ganymede, and calculated the resulting melt volume for every impactor that produces a crater of the size corresponding to the impact features that we have mapped. They then compared these volumes to pit volumes derived from our mapping and morphometry. They found that inferred melt volumes are, in general, much greater (typically one to two orders of magnitude) than our pit volumes, and suggested that a modest amount of melt drainage along with infiltration or re-freezing played a role in generating these features. In general, there is ample melt available to be mobilized for surface process and surface feature formation, and this has been explored by others, e.g. Elder et al. (2012) who proposed similar ideas and suggested that drainage of melt through impact-generated fractures could account for crater pits on Ganymede and other bodies in the Solar System. Caussi et al. (2024) built on the hypotheses of Elder et al. (2012) and Korycansky et al. (2022a,b) by proposing that a pool of melt beneath the center of an impact feature would be sealed by some of the water draining into fractures underneath it and then refreezing, while surface freezing of the melt pool would trap it as a subsurface lens of melt. The melt would cool, freeze, and expand over time, stressing the surrounding ice, which would eventually break the seal and lead to drainage of liquid into deeper cavities in the subsurface and subsequent roof collapse over the lens to form the pit at the center of the crater.

In accordance with these studies, we argue that the degree to which melt has drained into the subsurface is a crucial factor in determining the differences in pit configuration between pit craters. The single pit craters Lugalmeslam and Isis display higher pit volumes than the multiple pit craters Achelous and Tindr by a factor of at least 4. Efficient draining of a large volume of melt (relative to the size of the impact) beneath Lugalmeslam and Isis may account for their single, high volume pits, while smaller, disconnected volumes of melt may have been generated for Achelous and Tindr, the drainage of which formed irregular, small pits isolated from each other. While the development of several small volume pits versus a single, large volume pit does not seem to be dependent on crater diameter, the degree to which it is dependent on the age of the crater or the age of the target surface is also not altogether obvious based on the crater statistics (see Table 3) and geologic contexts (see Table 1) of the four pit craters that we have mapped. Tindr and Achelous, the two craters with multiple small pits, are both quite young but are located within older, dark, cratered terrain and younger, bright grooved terrain respectively. Lugalmeslam is an ancient single-pit crater that has impacted into dark, cratered terrain, while Isis is a single-pit crater that is probably young and has impacted into bright, grooved terrain.

If the pits represent locations of impact melt drainage into the subsurface, then we interpret the positive topographic relief of the annuli, which commonly display a hummocky, hilly morphology, to represent bulging of the surface resulting from the refreezing and expansion of the ponded melt lens prior to the breaking of the seal and the draining of the melt into the deeper subsurface, as hypothesized by Caussi et al. (2024). The simulations of Korycansky et al. (2022a) show that the final configuration of melt underneath impact features in the pit crater, dome crater, and anomalous dome crater classes forms a wide, shallow lens extending to approximately half the diameter of the crater itself, with a narrower column of melt extending deeper under the center of the crater. We find that the mean ratio of annulus to crater diameter is 55%, which corresponds well to the width of the shallow melt lenses that are formed in the simulations. Mean annulus relief above the crater floor ranges between 50 and 250 m, meaning that if this relief is due to the ~9% volumetric expansion caused by the freezing of a subsurface melt lens, then such a lens may range in thickness between 550 m and 2750 m (likely to be minimum values as some of the melt in the lens may also drain upon formation of the pit). Annuli are highly variable not only in terms of the ratio of their volumes to those of the pits (ranging between 0.01 and 13.79 across the four pit craters, which also happens to be the full range across all pit, dome, and anomalous dome craters), but also the ratio of their diameters to those of the pits (which exhibits a decreasing trend from pit craters to anomalous dome craters, with pit craters, dome craters, and anomalous dome craters showing mean ratios of 2.6, 2.0, and 1.4 respectively). This implies high variability in the degree to which melt remains frozen in the shallow surface versus drains as a liquid into the subsurface, and also that the melt lens becomes less wide relative to the column with increasing crater diameter.

Based on our observations, there seems to be little dependence of pit crater morphology (as expressed in the pit and annulus configuration) on crater size, age, or geologic context, indicating that these impacts are only penetrating to shallow depths within a rigid ice layer, and so their final morphology is not subject to variations in the thickness of an ice shell overlying a significantly warmer ice or liquid layer the way the morphologies of much larger impact features are. Since the physical state of the shallow subsurface that these smaller impact penetrate to is essentially the same for all geological contexts at all times in the satellite's history, the morphologies of pit craters are likely not explained by the hypotheses in Figures 2d and 2e. Instead, the hypothesis in Figure 2c, in which an isolated pocket of melt is generated by the impact, the volume and configuration of which depends more on the velocity and angle of impact than the subsurface conditions at the time of impact. In turn, the volume and configuration of the melt pocket will influence the degree to which impact melt drains versus remains and refreezes, and therefore the final form of the pit crater.

*Dome craters:* The three dome craters that we have studied all exist on Ganymede, are all between 100 and 110 km in diameter, and are all relatively youthful based on their crater statistics and well-preserved morphologies. Osiris is one of the youngest large impact craters on Ganymede, showing prominent, very bright rays that extend more than 1000 km from the center of the crater, and very high overall topographic relief, featuring the highest rim of any of our measured impact features (exceeding 2 km) as well as the largest crater volume (in fact, all three dome craters have the largest volumes of all impact features). We have also not mapped any craters larger than the minimum resolvable diameter of 5.46 km that superpose the crater or the ejecta in its immediate vicinity. Melkart also shows a bright ejecta blanket (albeit not nearly as bright as that of Osiris) and a low density of mostly <3 km diameter craters superposed on it and its immediate ejecta, giving it the lowest-plotting R-plot of any cratered impact feature we have measured on Ganymede, approximately contemporaneous with Gilgamesh. Melkart is exceptional in terms of the prominence of its dome and its annulus, both the highest of any that we have measured, while its annulus volume is second only to that of the much larger diameter Serapis. Its dome has essentially the same volume as the pit in which it is contained. Its facies also show unusual diameter ratios, with the rim being only ~10 km wide, making it one of the narrowest rims relative to the size of the impact feature (<10% of Melkart's radius), and the annulus being especially wide (~73% the radius of the impact feature). Our mapping of Melkart, the only impact feature with a central dome that contacts the annulus, shows that its dome crosscuts the annulus, confirming that the annulus is established prior to dome formation. Eshmun shows only three superposing craters, all about 4 km across. Melkart is situated within dark, cratered terrain, and Eshmun, while located only 440 km away from Melkart, superposes bright, grooved terrain of Sippar Sulcus. Osiris superposes the grooved terrain of Mumu Sulci.

The important morphological distinctions between the pit and dome crater morphological classes are that dome craters always feature single, large pits on their floors, and these pits always harbor a steep-sided, flat-topped dome. It appears that pit diameter and the pit/crater rim diameter ratio must reach threshold values in order for dome formation to occur: the red plots in Figures 4a and 4b indicate a transition diameter of ~25 km and a transition ratio of ~0.25 (with some overlap, the pit crater Lugalmeslam has a ratio of 0.28 and Eshmun, the smallest dome crater, a ratio of 0.25). The universality of single, large pits within the dome and anomalous dome crater morphological classes indicates that the phenomenon causing multiple small pits to form within the annulus in some pit craters is never replicated in the larger classes. We interpret this to be due to impacts large enough to produce craters in the dome morphological class and those larger invariably producing sufficient quantities of melt to result in the formation of a single, large pocket of impact melt that drains efficiently into the subsurface – the pit craters with single pits, Lugalmeslam and Isis, have pit volumes of 46 and 36 $km^3$ respectively, which the pit volumes of dome and anomalous dome craters exceed by a factor of at least 4.5. As with the pit craters, we do not interpret the pre-existing regional geologic context of dome craters to have much bearing on their final morphology. The fact that the dome craters arguably display the least variety in the configuration of their facies of any impact feature size class (always showing an outwards sequence of dome, single pit, annulus, floor, and rim with fairly consistent facies diameter ratios, with the exception of their annuli) indicates that impacts in this size class are the least sensitive to pre-existing geological context in terms of controlling their final crater morphology. It is harder for us to assess whether the timing of their formation is influential as all three of the dome craters we have examined are young. As with the pits, the prominent annuli of the dome craters, the volumes of which exceed those of the pit craters by at least a factor of 1.6, are a testament to the larger volumes of melt produced by the larger impacts that formed these craters, with an especially high fraction of the melt remaining at a shallow level in the subsurface and refreezing to form Melkart's very tall and wide annulus.

Caussi et al. (2024) have addressed the issue of dome formation within craters on Ganymede and Callisto by using finite element simulations to model relaxation of crater topography, in which they use our averaged topographic profiles of certain impact features in Figure 3 to evaluate their results. These same methods have been used to investigate crater relaxation on the Saturnian mid-sized icy satellites (White et al., 2013, 2017). The simulation results of Caussi et al. (2024) cause them to argue that dome craters can evolve from pit craters through topographic relaxation, facilitated by the remnant heat from the impact, with the domes forming within ~10 to ~100 Myr after impact. The results show that topographic relaxation acts to eliminate the stresses induced by the crater topography and restore a flat surface: ice flows downwards from the rim and upwards from the crater depression driven by gravity. Caussi et al. (2024) used a pit crater as starting topography for their simulations, which, as described in the previous section, is hypothesized to have formed via the partial freezing and subsequent drainage of a trapped melt pocket. Following this drainage, any remaining liquid melt would freeze, with remnant heat left over from the impact that is concentrated below the pit causing this ice to be relatively warm, and therefore to have reduced viscosity. Since this warm ice flows more readily than the colder ice under the crater floor that surrounds it, the upward flow is enhanced beneath the pit, leading to the emergence of a dome. Caussi et al. (2024) found that their simulation result that provided the closest match to our topographic profile of Osiris was obtained when the thermal anomaly was confined strictly below the pit. Notably, their simulations show that formation of a dome is always accompanied by relaxation of the crater as a whole, because enhanced uplift in the pit can only occur when the surrounding ice is soft enough to accommodate this extra flow. Longer wavelength topography relaxes more readily than shorter wavelength topography of the same amplitude, and dome formation is only viable for impact features in the dome crater size class and above. Simulations performed for craters in the pit crater size class (several tens of km diameter) show that while the crater can relax under a high heat flux, the pit sizes are too small to allow a dome to form within the pit via relaxation, regardless of heat flux. At the transition diameter between pit and dome craters, domes only emerge if the central pit is wide, and the heat flux is high.

Simulations performed by Caussi et al. (2024) for different heat fluxes show that higher fluxes lead to taller domes: those performed for a low heat flux of 3.5 mW $m^{-2}$ yielded a dome summit elevation that is comparable to that of the annulus crest, while those performed for a high heat flux of 10 mW $m^{-2}$ yielded a dome summit elevation that exceeds that of the rim crest, while uplift of the pit results in it being shallow relative to the crater floor. This is seen at Melkart, the dome summit of which is higher than the rim crest by ~200 m, while its pit is the shallowest (i.e. most uplifted) of all the dome craters at 210 m, indicating that this dome crater has experienced an elevated heat flux lasting at least ~10 Myr at some point after its formation. Given Melkart's youth, its very relaxed morphology shows that high heat flows did characterize Ganymede's relatively recent history, even if they were localized geographically and temporally. Eshmun and Osiris do not show similarly uplifted topography, indicating that they have only ever experienced lower heat fluxes. The simulations show that relaxation has reduced the floor depths of dome craters by as much as 45% and 65% over 100 Myr for heat fluxes of 3.5 mW $m^{-2}$ and 10 mW $m^{-2}$ respectively. This is reflected in our observation that the mean floor depth of the dome craters (0.29 km) is actually slightly less than that of the pit craters (0.31 km) despite the dome craters having larger diameters by tens of km.

The simulations of Caussi et al. (2024) show that the bulk morphologies of these domes can be plausibly explained by relaxation of pit topography, but the surface textures of the domes indicate that other factors may also contribute to their growth. A vaguely lineated "breadcrust" texture is identified on the better resolved domes, including those of Melkart, Har, Doh, and Neith, which may indicate fracturing of a rigid outer carapace that formed at their surfaces. Caussi et al. (2024) noted that some remaining liquid melt within the pocket beneath the pit might extrude through the collapsed roof of the pocket onto the floor of the pit, which they argued could "seed" an initial small dome that subsequently grows via relaxation; we consider it possible that ongoing extrusion of such melt might also supplement the relaxation and contribute to the expansion and fracturing of the dome's cooled surface. Caussi et al. (2024) also found that relaxation results in raised pit rims (i.e. what we have mapped as annuli) for dome craters under both high and low heat flow conditions, and for smaller pit craters under high heat flow conditions, but not for pit craters under low heat conditions, causing them to conclude that a formation mechanism besides relaxation is likely required for the formation of the annuli. As we have already noted, re-freezing and expansion of trapped melt surrounding the central zone of drainage under the crater is a viable alternative mechanism for uplifting the crater floor around the pit, although for larger craters topographic relaxation may supplement the relief of the annulus.

*Anomalous dome craters:* The two smallest of the five anomalous dome craters we have examined are located on Callisto: Har and Doh. While both of these craters are covered at least in part by high resolution imaging, assessing their morphologies is inhibited for different reasons. The center of Har is located only 115 km from the center of the younger pit crater Tindr, the ejecta and secondary craters of which have modified Har's appearance, and possibly obscured details of some of its facies. We interpret most of the 75 impact craters of 5 km diameter or less that are superposed on Har to be Tindr secondaries, as well as the crater chains that we have mapped. The high density of these secondaries place Har high on the R-plot, but the single 21.5 km diameter crater that is also superposed on it is not a secondary, and it alone indicates that Har is genuinely an ancient crater. Har has no obvious crater floor or rim surrounding its annulus (the topographic relief of which is the most muted compared to all other dome and anomalous dome craters), but rather an outer platform that is raised slightly above the surrounding terrain (this is the measurement shown in the rim height column for Har in Table 2). As we describe in section 4, this outer platform presumably constitutes both the floor and rim of Har, but which cannot be distinguished owing to topographic relief and morphologies of both having been obscured by Tindr ejecta. Doh, located near the high albedo center of the Asgard multi-ring feature, is covered by the highest resolution imaging of any impact feature in our study, but this only covers the impact feature as far out as its annulus in its entirety. While high resolution imaging does extend far beyond the annulus to the south, as well as a short distance beyond the annulus in other directions, and while the terrain immediately surrounding the annulus must logically represent the floor of the crater, no contact of this floor with a rim can be confidently identified anywhere in either the high resolution imaging or the global mosaic beyond. As such, Doh remains unmapped beyond the annulus and we have inferred its rim diameter of 128 km based on the linear relationships of pit and annulus diameter to impact feature diameter as shown by the relevant trend lines in Figure 4a. The portion of Doh that we have mapped is lightly cratered and plots between Tindr and Har on the R-plot, roughly contemporaneous with Valhalla. A consequence of these limitations is that neither floor depths nor crater volumes have been measured for either Har or Doh.

Of all the anomalous dome craters, Neith on Ganymede is both the best preserved and that which has the highest fraction of its expanse covered by high-resolution imaging. This means that, unlike either Har or Doh, its floor and rim are both distinguishable in the imaging and DEM. Like Doh, Neith's superimposed craters mostly have diameters of a few km, with one larger than 10 km, giving it a middling position on the R-plot. Neith is situated in dark, cratered terrain just north of bright, grooved terrain of Phrygia Sulcus. Anzu and Serapis, the largest anomalous dome craters, were imaged at only 2 km/pixel across their whole expanses, with the imaging for Anzu being quite oblique. Both, however, were imaged at a low Sun angle, accentuating their topography, and robust photoclinometric DEMs were produced for both of them that allow reasonably confident mapping of their facies. No superimposed impact craters of 14 km diameter and above are resolved for Anzu, while only two are resolved for Serapis, making it plot fairly high on the R-plot. The inability to identify small craters like those superposing Har, Doh, and Neith, however, means that the ages of Anzu and Serapis cannot be constrained to the same degree based on crater statistics. Both Serapis and Anzu superpose bright, grooved terrain.

Anomalous dome craters broadly display the same facies as dome craters. As with dome craters, anomalous dome craters display prominent annuli, pits, and domes, with the main characteristic that distinguishes them being that their floors and rims show markedly less topographic relief and are less well defined than those of dome craters. This is best seen at the well-resolved Neith, which displays the deepest pit of any impact feature (0.7 km) and an annulus and dome that are amongst the highest measured, but which shows a highly muted rim that is essentially a broad, irregular annulus varying between 3 and 40 km wide, is raised above the floor by only 1 km, and which displays no crest. A similar pattern is seen for Anzu and Serapis, despite the lower resolution imaging that covers them. In fact, the low topographic relief of the outer portions of these impact features is highlighted by the fact that the mean elevations of their floors actually exceed that of their surroundings by 50 to 100 m (a consequence of which is that we have not calculated crater volumes for either of these impact features. Unlike the rims of Neith and Anzu, the rim of Serapis appears to be only partial. One way in which at least some anomalous dome craters appear to be more complex morphologically that the dome craters is the occasional presence of a ~500 m deep, partial bounding trough beyond the rim, which is seen at Neith and Anzu. This may be representative of a transition to palimpsest-like impact features, in which the outer bounds of the undulating plains are sometimes seen to dip down by a few hundred meters, like along the northern boundary of Hathor.

Even though their annuli, pits, and domes are topographically prominent in comparison to their floors and rims, there is still a general tapering of the relief of these facies with increasing diameter of the anomalous dome craters as shown in Figure 4: pit depth and annulus height increase as far as Neith, decreasing thereafter, while dome height peaks at Anzu, decreasing for Serapis. Similarly, pit volume and dome volume peak at Neith and Anzu, and decrease for Serapis, although annulus volume continues to increase as far as Serapis, indicating that any decrease in annulus relief is not compensated for by a corresponding decrease in area for the largest anomalous dome craters. Caussi et al. (2024) performed relaxation simulations for a 145 km diameter crater in the anomalous dome class and used our averaged topographic profile of Neith in Figure 3 as the means to evaluate it. Note that our 170 km diameter for Neith refers to the distance to the outer bound of the topographically muted rim facies, while their simulated diameter of 145 km corresponds to the distance to the topographic rise in the middle of the rim as it appears in our averaged topographic profile, which is where a sharp rim crest separating the inner and outer walls of the rim would be expected for a dome or pit crater. For a low heat flux of 3 mW $m^{-2}$, a prominent dome rising ~1.1 km above the pit floor emerges in the simulation (very similar to the 1.15 km measured for the Neith dome), with the final crater rim showing relief above the floor of ~1 km, which matches the 1 km measured for Neith, but which also shows a sharper crest than Neith's rim does. For the high heat flux of 10 mW $m^{-2}$, all topography is relaxed within 10 Myr (with total crater relief of ~100 m). Such relaxed topography is not reflected in any of the anomalous dome craters that we have measured, indicating that none of them have been subjected to heat fluxes this high at any time in their histories. This appears to be consistent with our observation that the majority of the anomalous dome craters we have studied (with the exception of Har) are middle-aged or younger based on their crater statistics and/or their superposition of bright, grooved terrain, and therefore that they formed after the high heat fluxes that prevailed during the early histories of these satellites had subsided.

The subdued topographic relief of the floors and rims of anomalous dome craters represents the increasing influence of the physical state of the subsurface with increasing impact feature diameter, whereby impacts that form anomalous dome craters and penepalimpsests/palimpsests are large enough to penetrate into a warm subsurface zone, which for the younger anomalous dome craters would likely entail warm subsurface ice (as in Figure 2d), rather than a liquid layer underneath warm subsurface ice (as in Figure 2e). Relaxation of long-wavelength topography also contributes to the subdued floor and rim relief, as demonstrated in the simulations of Caussi et al. (2024) for impact features in the anomalous dome crater size range, but is not sufficient to eliminate the short-wavelength topography of sharp crater rims (Fig. 10 in Caussi et al., 2024), meaning that the rims were initially low and broad. The penetration of the impact into a weaker subsurface layer would cause the material that is mobilized to define the initial crater shape to behave more like a Bingham viscoplastic (i.e. behaving as a rigid body at low stress but flowing as a viscous fluid at high stress) than if only the more rigid shallow subsurface layers had been penetrated as in the case of a smaller impact. This more plastic rheology means that the initial crater cannot support the same high topographic relief as for the smaller dome and pit craters, hence the decreasing floor depths and rim heights with increasing impact feature diameter as shown in Figure 4c. However, the fact that the annuli and pits of anomalous dome craters are still often topographically prominent (especially in the case of Neith) shows that, while the initial crater presents minimal topographic relief, the impact process is still accompanied by the production of large volumes of melt, some of which drains to form deep pits, and some of which remains in the shallow subsurface and refreezes to form the annulus. The melt volume calculations of Korycansky et al. (2022b) determined that, as with the pit and dome craters, the pit volumes of anomalous dome craters tend to be smaller than the melt volumes produced by impacts large enough to form such craters by a factor of 1-2 orders of magnitude, implying that only a minority of the produced melt contributes to the observed annulus and pit morphology.

*Possible transitional features between craters and penepalimpsests:* The largest impact feature class that we examine, comprising penepalimpsests and palimpsests, is unique in that nearly all of the facies that define it (excepting the dome that we have mapped for the raised central portion of Hathor) are unique to this class, exemplifying the profound morphological transition from anomalous dome craters. Instead, they show a common configuration of a small central expanse of smooth plains bounded by low scarps (again, excepting Hathor's raised central plains) that is surrounded by a broad expanse of undulating plains that stretches all the way to the edge of the impact feature, within which are distributed varying concentrations of concentric ridges. The contact of the undulating plains with the surrounding terrain is often subtle, with no distinct topographic rim, and is generally indicated by a transition in surface texture, best seen in the DEM when available. We note that the term penepalimpsest has been used somewhat flexibly in the literature, sometimes to describe features that are interpreted to be transitional between anomalous dome craters and palimpsests (Passey and Shoemaker (1982) even described what we call anomalous dome craters as "Type II penepalimpsests"), while other studies, such as Thomas and Squyres (1990) and Jones et al. (2003) have lumped penepalimpsests and palimpsests together. Schenk et al. (2004) describe several concentric structures in penepalimpsest interiors, usually inward-facing scarps or narrow low ridges several hundred meters high, in addition to the central smooth area. We do not place penepalimpsests and palimpsests in separate impact feature classes, but rather regard penepalimpsests as being those impact features within this single class (specifically Hathor, Buto Facula, and Nidaba) that display a higher proportion of concentric ridges within their interior, which also reach greater lengths, extend around a greater fraction of the impact feature's circumference, and appear more developed than those in impact features that are better described as palimpsests (specifically Teshub, Zakar, Epigeus, and Memphis Facula).

The large impact features of Ganymede and Callisto therefore essentially fall into two broad morphological classes: craters, where the impact forms a depression surrounded by a rim, and the penepalimpsests and palimpsests, where the impact does not form a depression, and the overall topographic relief of the feature is minimal (few hundred meters). The morphological disparity between pit and dome craters is essentially governed by diameter, with domes emerging on pit floors above a threshold crater diameter. For impact features larger than dome craters, morphology appears to be less size-dependent: while most of our measured impact features between 110 and 200 km in diameter are anomalous dome craters, the smaller penepalimpsests (the smallest being Hathor at 173 km diameter) and palimpsests also occur in this size range, while Serapis, the largest anomalous dome crater, is 253 km in diameter. This observation demonstrates that size becomes a less crucial factor in determining impact feature morphology above diameters of 150 km, with the physical state of the subsurface (which itself may be related to the timing of the impact) becoming increasingly important. We interpret the morphologies of anomalous dome craters to indicate that large impact features in this size class are beginning to demonstrate sensitivity to the effects of a warm, weak, but still solid subsurface layer, but which are still also exhibiting the effects of subsequent re-freezing and drainage of impact melt beneath the crater.

Within this diameter range in which the morphologies of large impact features appear to be less size-dependent, there are some impact features with morphologies that appear transitional between anomalous dome craters and penepalimpsests. Comparison of the morphologies of Serapis and Hathor is revealing for understanding this transition. The essential facies that form anomalous dome craters can be recognized in Serapis, but its annulus and rim have dissociated into a fragmented, concentric configuration. For Hathor, this transition has advanced such that it presents a set of concentric ridges surrounding an expanse of smooth plains at the center. The majority of the impact feature consists of undulating plains, with no identifiable rim, crater floor, or central pit. Interestingly, both of these features display what we have mapped as the dome facies at their center, although they may be more accurately described as raised central plains, since both feature the smallest height/diameter ratios of all domes that we have mapped (0.012 and 0.016 for Serapis and Hathor respectively compared to a mean of 0.028 across all domes). Hathor is the smallest feature in the penepalimpsest/palimpsest feature class that we have studied, and also the only one that shows such a dome, which is elevated by ~400 m above the undulating plains that surrounds it and which it superposes. Unlike the dome of Serapis, which is located within a broad central pit, as is characteristic of anomalous dome craters, the dome of Hathor is not situated within a pit. As such, while the dome of Serapis can feasibly form via gravity-driven flow of warm, low-viscosity ice, the driving force of relaxation (Caussi et al., 2024), such a formation mechanism is not viable for Hathor's dome given that it is not present within a topographic low. Rather, we consider Hathor's dome to more likely represent an extrusion of material onto the surface, as with the central plains facies (see following section), but which is more viscous and so forms a scarp-bounded edifice rather than an expanse of plains.

The factors that determine why these impact features both display central domes are the same that govern why there is an overlap in size of the largest anomalous dome craters and the smallest penepalimpsests and palimpsests between 170 and 250 km diameter. Serapis has a diameter ~1.5 times that of Hathor and so its impact might be expected to be more energetic and to liquefy a higher proportion of the target material than Hathor's, producing either a comparable penepalimpsest or even a palimpsest. However, we argue that the Serapis impact penetrated into a warm, viscous subsurface ice layer (like that in the scenario in Figure 2d), which was sufficiently rigid such that liquefaction of the target material was limited and the resulting impact structure organized into the radial sequence of somewhat fragmented but still recognizable facies, each with distinctive topographic relief relative to those neighboring it, causing it to form an anomalous dome crater. Due to the low resolution of the imaging covering Serapis, its age cannot be constrained terribly well based on crater statistics, but its impact into bright, grooved terrain indicates that it is relatively young and so impacted at a time in Ganymede's history when heat fluxes were lower and its lithosphere was generally thicker across the globe, with fewer instances of near surface liquid layers. As we have noted in the previous section, all the anomalous dome craters we have mapped (with the exception of Har) appear to be relatively young, based either on crater statistics or them superposing young grooved terrain, and so it is more likely that they have impacted into a thick, rigid lithosphere. In the case of Hathor, while the energy of its impact was smaller than that of Serapis, we argue that the physical state of the subsurface was more akin to that in the scenario in Figure 2e, whereby the impact penetrated into a pre-existing subsurface liquid layer, resulting in liberation of a substantial volume of this liquid, and precluding the organization of the impact structure into the ordered facies of an anomalous dome crater, with the low, concentric ridges distributed amongst the undulating plains being the only positive relief that the structure could support. The subsequent extrusion of a dome at the center of the impact feature is likely a consequence of the fact that the impact energy associated with the Hathor impact was relatively small by the standards of penepalimpsests, with relatively little remnant impact heat being retained underneath the impact feature, which would cause the extruded material to have a higher viscosity than the material forming the central plains facies of larger penepalimpsests and palimpsests.

Since it is covered by higher quality imaging than Serapis, our crater statistics for Hathor are more robust, and it does plot slightly higher than Serapis on the R-plot (certainly for the 19 km crater diameter bin that they share). Hathor is located on the periphery of Bubastis Sulci, a few of the grooves of which it crosscuts, but its more cratered appearance relative to Serapis indicates that it is the older feature, and impacted at a location on Ganymede that experienced a higher heat flux that permitted a subsurface liquid layer. The other penepalimpsests that we have studied, Nidaba and Buto Facula, are both located within dark, cratered terrain and plot similarly to Hathor on the R-plot, with all three nearly overlapping across the 9.5 to 22.6 km crater diameter range that they share, so are likely to be of similar age. Schenk et al. (2004) noted that the superposed crater densities of penepalimpsests are very similar to those of bright terrain, suggesting that they are either coincident with or just postdate the bright terrain, which our observations and crater statistics support.

*Penepalimpsests/Palimpsests:* As with anomalous dome craters and penepalimpsests, there is not a definite cutoff diameter separating penepalimpsests from palimpsests. We have described in the previous section how penepalimpsests are characterized by relatively prominent and closely spaced ranges of concentric ridges. In contrast, palimpsests display sporadic ranges of ridges that show a less obvious concentric fabric, and so possess even less positive topographic relief than the penepalimpsests. The largest impact feature that we have mapped, the 354 km wide Memphis Facula, is also the definitive palimpsest, displaying no concentric ridges at all, and instead appearing as an albedo feature, with much of its expanse mapped as undulating plains of varying brightness. Our mapping shows that concentric ridges occupy a mean fraction of 13% of the total areas of the features that we have classified as penepalimpsests, while they occupy only 6% of the areas of the features that we have classified as palimpsests.

Schenk et al. (2004) regarded the smooth central plains of penepalimpsests and palimpsests as being the equivalent of central domes, and our mapping of this facies shows that they do superpose, and so postdate, the undulating plains that surround them, like the domes of the smaller impact feature classes. While penepalimpsests and palimpsests do not display pits, their central plains are bounded by low scarps separating them from the undulating plains, giving the central plains the appearance of a low viscosity flow that has spread across a depression at the center of the impact feature. They therefore arguably might represent late emplacements of viscous material into central depressions that are no longer apparent due to their volumes having been entirely occupied by this material. The very flat surface of these central plains and their lack of positive relief suggest a state of hydrostatic equilibrium, and cause us to suggest that emplacement proceeded via extrusion of a low-viscosity flow that filled the central region of the impact feature, rather than via relaxation of ice under the depression floor that was warmed by remnant impact heat, as Caussi et al. (2024) have hypothesized for dome formation. The low-viscosity material may be a combination of impact melt and pre-existing subsurface material, either fluidized by the impact or sourced from a pre-existing fluid layer, which these impacts have plausibly penetrated owing to their large size. Unlike the ~10 Myr that the simulations of Caussi et al. (2024) show is required for a dome to develop via relaxation in dome and anomalous dome craters, the infilling of any depression that is present at the center of a penepalimpsest/palimpsest with extruded, low-viscosity material would more likely occur in the incipient development stage, immediately following the impact.

As we have noted, penepalimpsests and palimpsests essentially represent different ends of the morphological spectrum of this single impact feature class, with concentric ridges forming a higher fraction of penepalimpsests than palimpsests. Whether an impact forms a penepalimpsest or a palimpsest is largely dependent on the same factor that governs whether a penepalimpsest or an anomalous dome crater forms, i.e. the physical state of the subsurface at the time of impact. We interpret formation of an anomalous dome crater versus a penepalimpsest to be dependent on whether the impact is penetrating a subsurface layer of warm ice versus a fluid layer, respectively. For penepalimpsests and palimpsests, we interpret both to be impacting into a subsurface fluid layer, with the morphological difference between the two (whereby penepalimpsests display more positive relief topography than palimpsests) hinging on the thickness of the overlying ice layer and therefore the degree of liquidization of the target material and how long this material persists in the liquid state. The nearly flat appearance of palimpsests, which are mostly occupied by the undulating plains, and the total absence of a crater rim, suggests that the impacts that create these features involve the generation of a vast quantity of low-viscosity fluid that mostly consists of pre-existing fluid in a shallow subsurface layer that has been liberated by the impact, with only a minor element being solid material of the bolide and the target that has been melted by the impact. Rather than forming a crater with a raised rim that is surrounded by ejecta, the impact would instead fluidize the target surface out to a certain radius from the impact zone, particularly evident at Buto (Schenk et al., 2004; Moore et al., 2024). The efficiency of this fluidization, and therefore whether a penepalimpsest or a palimpsest forms, would depend on the size of the impact and the thickness of the solid ice layer overlying the subsurface fluid layer: a smaller impact and a thicker ice layer would result in less efficient fluidization, with the consequence that a higher proportion of solid material survives the impact process to be uplifted to form the concentric ridges.

Schenk et al. (2004) described palimpsests as having higher crater densities than penepalimpsests, and occurring only on older terrains while penepalimpsests occur on all terrain types, and so argued that the difference between the two is primarily related to age. It is logical that palimpsest formation would be more common during the early histories of these satellites, when the global heat flux was at its highest, and a subsurface liquid layer extended to generally shallower depths than was possible later. This would mean that an impact of a given size would be more likely to form a palimpsest in a thin ice shell earlier in a satellite’s history, and more likely to form a penepalimpsest in a thicker ice shell later in its history. Our crater statistics show that two palimpsests, Teshub and Memphis Facula, are amongst the most ancient of all of the impact features we have examined, displaying high R-plots especially in the 5 to 14 km crater diameter range. These two palimpsests also display the lowest proportion of concentric ridges within their interiors, indicating the ubiquity of low-viscosity fluid in defining the final morphologies of these features, with virtually no solid element being uplifted to produce the ridges. Hathor and Teshub represent a compelling case of the influence of the timing of impact on whether a penepalimpsest or a palimpsest is formed. These impact features are of comparable size (Hathor's diameter is 173 km, Teshub's is 188 km) and they are located proximal to each other, with their centers being only 200 km apart. Teshub, however, is very much a palimpsest, with very little positive topographic relief, and is very ancient as indicated not just by its high crater count, but also by the fact that its western portion has been crosscut and obliterated by the later forming Bubastis Sulci. Hathor stands in contrast by showing abundant and fairly well ordered concentric ridges, as well as its central dome rather than central plains like Teshub has. It shows a lower crater count than Teshub, with its R-plot being lower than Teshub's across all crater diameters, and as we have noted earlier, it also crosscuts some of the fractures of Bubastis Sulci. So while these two similarly sized impact features have formed close to each other, the interval between the timing of their impacts, during which the heat flux in this part of Ganymede declined and a subsurface liquid layer deepened below a thickening ice layer, was the crucial factor in determining their different appearances.

The assertion that palimpsests are old and penepalimpsests are young is a generalization, and impactor size is still an important factor in determining impact feature morphology, albeit not to the extent that it is for smaller impact feature classes. Teshub and Memphis Facula are reasonably well resolved in the imaging that covers them, and we are confident in our crater statistics that mark them out as being ancient. The other palimpsests we have studied, Zakar and Epigeus, are not as well resolved, and while they both plot lower than all the other palimpsests and penepalimpsests on the R-plot, we cannot constrain their ages as well based on crater statistics. However, both Zakar and Epigeus superpose bright, grooved terrain, while Memphis Facula exists within dark, cratered terrain and Teshub is crosscut by bright, grooved terrain, so the geologic contexts of Zakar and Epigeus do also support them being young palimpsests. The palimpsests we have mapped are generally larger than the penepalimpsests, displaying a mean diameter of 289 km as opposed to 224 km for penepalimpsests, and for a given ice shell thickness over a pre-existing liquid layer, a larger impact will succeed in liberating a larger volume of liquid from this layer than a smaller one, meaning that even during quite late stages in Ganymede's history when the ice shell has thickened, a very large impact is still sufficient to produce a palimpsest, even if smaller impacts can only yield anomalous dome craters and penepalimpsests. It is unsurprising that Memphis Facula is the only palimpsest for which we have mapped no concentric ridges, given that it is the largest of all the impact features we have studied with a diameter of 354 km, and is also very ancient based on displaying a high crater count and being located within dark, cratered terrain, meaning that its very large impact into a very thin ice shell overlying a liquid layer would have released the highest volume of liquid of any penepalimpsest or palimpsest.

## 7. Conclusions

Our geologic mapping, morphometry, and crater age dating study of a variety of large impact features on Ganymede and Callisto, when considered alongside parallel modeling studies that have used our measurements to evaluate their simulations of the volume of impact melt generated by the impacts as well as their subsequent modification by viscous relaxation, has cast additional light on the longstanding question of why these features are so morphologically diverse. By creating DEMs for nearly all the impact features under investigation and using them to gather morphometric statistics relating to the common facies that make up each impact feature class, tracing how these statistics vary between classes, and estimating the relative ages of the impact features based on their superposed crater densities and geologic contexts, we have identified what factors are instrumental in determining the distinctive morphologies of each of the impact feature classes. We have found that certain hypotheses that have been formulated to explain these features, which we outlined in the introduction, may apply to a particular impact feature class, while another hypotheses is more appropriate for explaining a different impact feature class. The large impact impact features of these satellites broadly fall into two morphological classes: craters (including pit, dome, and anomalous dome craters) and penepalimpsests/palimpsests, although at the transition between these two main classes we have identified a handful of features (including Hathor and Serapis) that display characteristics of both and do not obviously fall into either one.

The pit volumes that we have measured for all impact feature classes are always far exceeded by the melt volumes generated by the simulations of Korycansky et al. (2022b), indicating that subsequent mobilization and evolution of only a small minority of this melt is reflected in the impact feature morphology. As other studies have hypothesized (Elder et al., 2012; Korycansky et al., 2022b; Caussi et al., 2022) the majority of the melt drains into the subsurface through fractures, causing surface collapse at the center of the crater to form a pit, while we interpret the raised annuli that surround the pits to be evidence that some of the melt refreezes and expands to raise the crater floor surrounding the pit. Pit craters display either a single, high volume pit, or multiple, much smaller volume pits, which we interpret to reflect the degree of melt drainage under the crater versus melt remaining and refreezing to form the annulus, but there appears to be little correlation of these different morphologies with crater diameter, age, or geologic context, implying that the velocity and angle of impact may play a prominent role in determining the volume and configuration of the pits. We also do not identify any correlation of the morphologies of pit craters and dome craters with age or geologic context. Whether a pit crater or a dome crater forms is essentially a matter of impact size, whereby craters with diameters in the vicinity of 100 km possess sufficiently large central pits such that they experience subsequent viscous relaxation of their pit topography (aided by remnant impact heat) that forms a dome on the floor of the pit, as shown for a range of heat fluxes in the simulations of Caussi et al. (2022). Our results indicate that neither pit nor dome craters have penetrated deep enough to reach a weak subsurface layer, which may be either warm ice or liquid water, at any time in the histories of these satellites, and therefore that their morphologies are never subject to variations in the physical state of the subsurface, which has varied over time as the surface ice shell thickens above a warmer, weaker, more fluid layer. Rather, the morphologies of pit and dome craters appear to be dependent on the volume and configuration of the subsurface melt pocket that is generated by the impact, as well as the amount of remnant impact heat generated by the impact, both of which largely depend on the size of the impact. As such, we argue that the scenarios in Figure 2b and 2c are those that most likely pertain to pit and dome craters.

The morphological transition from dome to anomalous dome craters reflects the emergence of the physical state of the subsurface as an influence on impact feature morphology. Compared to pit and dome craters, anomalous dome craters display shallow crater floors and low, subdued crater rims, causing us to interpret the large impacts that produced these craters as having succeeded in penetrating to a warm ice layer, causing the resulting impact structure to be less able to support high topographic relief compared to smaller craters that only penetrated into colder, rigid ice, while also facilitating subsequent relaxation of the floor and rim. However, anomalous dome craters still display quite prominent pits and annuli, indicating that generation of impact melt and its subsequent drainage and refreezing is still an important factor in determining the final morphologies of these craters.

A consequence of large impactors penetrating a weak subsurface layer is that impactor size, while still influential, becomes a less important factor in determining impact feature morphology. The morphological transition from pit to dome to anomalous dome craters is essentially size-dependent, but for impact features with diameters above ~170 km, whether an anomalous dome crater, a penepalimpsest, or a palimpsest forms is dependent on whether there is either a warm ice or else a liquid near-subsurface layer present, and how shallow and thick that layer is, as well as the size of the impactor. Since the physical state of the subsurface across these icy satellites has evolved across their histories, this also means that the timing of these large impacts is more influential on their morphologies compared to smaller impacts. For penepalimpsests and palimpsests, which both display very minimal topography relief, mobilization of a pre-existing subsurface liquid layer is integral to determining their morphology, with the fate of any melt generated by the impact being minor in importance in this respect. A more viscous rheology of the mobilized material appears to have characterized the impacts forming penepalimpsests owing to concentric ridges composing a larger fraction of them relative to palimpsests. The commonality of the concentric ridges might reflect the thickness of the ice layer above the liquid layer, with the absence of ridges meaning that the surface ice layer was thin at the time of impact. Anomalous dome craters are the smallest of these three large impact feature morphology types, ranging in diameter between 110 and 253 km, and tend to be relatively young. This causes us to interpret them as having been formed by quite late impacts into a subsurface layer of warm ice rather than liquid (subscribing to the scenario in Figure 2d), with the impact not being large enough to produce vast quantities of liquid melt that are comparable in volume to the liberated quantities of pre-existing, near-subsurface liquid that are necessary to form a penepalimpsest or palimpsest, with the resulting impact structure still retaining the essential features of an impact crater. The penepalimpsests are the second largest type, ranging in diameter from 177 to 265 km, and are generally younger than palimpsests, the largest type, which range in diameter from 188 to 354 km. The very large impacts forming palimpsests will therefore also tend to penetrate through a thinner ice shell overlying the liquid earlier in the satellite's history, generating and liberating a larger volume of liquid than the smaller impacts forming penepalimpsests, which tend to form later in the satellite's history, once a reduction in heat flux has caused the overlying ice shell to thicken. We discuss and explain the appearances of penepalimpsests and palimpsests in detail in Moore et al. (2024).

This study was made possible based primarily on the availability of Galileo stereo imaging across Ganymede and Callisto, which covers only a small portion of their surfaces, and as such our survey considers only a small fraction of the large impact features that have been identified on these two satellites. Consequentially, our view of the total range of large impact feature morphologies on these two satellites, including those that may be transitional between the main morphological classes that have been defined, remains far from complete. The data returned by the Jupiter Icy Moons Explorer (JUICE), which will perform multiple flybys of Callisto before entering orbit around Ganymede in the 2030s, will greatly supersede Galileo's. The entirety of Ganymede and portions of Callisto are expected to be imaged at better than 400 m/pixel, with selected targets being investigated at better than 25 m/pixel, while a laser altimeter with a 20 m spot size and 10 cm vertical precision will be deployed upon orbital insertion around Ganymede. In addition, Europa Clipper will make approximately a dozen close flybys of Ganymede and Callisto and potentially acquire complementary observations. JUICE and Europa Clipper imaging, topography, and ground penetrating radar sounding will greatly expand the list of well resolved large impact features on these two satellites, and will permit a similarly expanded investigation combining mapping, morphometry, subsurface sounding, and modeling that will further refine what conditions are instrumental in governing impact feature morphology.

## Acknowledgements

Funding for this investigation was provided by the NASA Solar System Workings Program award 80NSSC19K0551.